\documentclass[%
 reprint,
superscriptaddress,
 amsmath,amssymb,
 aps,
 floatfix,
]{revtex4-2}

\usepackage{graphicx}
\usepackage{dcolumn}
\usepackage{bm}
\usepackage{siunitx}
\usepackage{amsmath}

\begin{document}

\preprint{APS/123-QED}

\title{Nanoscale Mapping of Magnetic Orientations with Complex X-ray Magnetic Linear Dichroism}

\author{Marina Raboni Ferreira}
\affiliation{Max Planck Institute for Chemical Physics of Solids, 01187 Dresden, Germany}
\affiliation{``Gleb Wataghin'' Institute of Physics, University of Campinas, 13083-859 Campinas, Brazil}
\affiliation{Brazilian Synchrotron Light Laboratory, Brazilian Center for Research in Energy and Materials, 13085-970 Campinas, Brazil}
 \email{marina.ferreira@lnls.br}
\author{Benedikt J. Daurer}
\affiliation{Diamond Light Source, Harwell Science and Innovation Campus, Didcot 0X11 ODE, United Kingdom}

\author{Jeffrey Neethirajan}
 \affiliation{Max Planck Institute for Chemical Physics of Solids, 01187 Dresden, Germany}

\author{Andreas Apseros}
\affiliation{Laboratory for Mesoscopic Systems, Department of Materials, ETH Zurich, 8093 Zurich, Switzerland}
\affiliation{PSI Center for Neutron and Muon Sciences, 5232 Villigen PSI, Switzerland}
\author{Sandra Ruiz-Gómez}
\affiliation{Max Planck Institute for Chemical Physics of Solids, 01187 Dresden, Germany}%
\affiliation{ALBA Synchrotron Light Source, 08290, Cerdanyola del Valles, Barcelona, Spain}

\author{Burkhard Kaulich}
\affiliation{Diamond Light Source, Harwell Science and Innovation Campus, Didcot 0X11 ODE, United Kingdom}%
\author{Majid Kazemian}
\affiliation{Diamond Light Source, Harwell Science and Innovation Campus, Didcot 0X11 ODE, United Kingdom}%
\author{Claire Donnelly}
\affiliation{Max Planck Institute for Chemical Physics of Solids, 01187 Dresden, Germany}%
\affiliation{International Institute for Sustainability with Knotted Chiral Meta Matter (WPI-SKCM2), Hiroshima University, Hiroshima 739-8526, Japan}
\email{claire.donnelly@cpfs.mpg.de}

\begin{abstract}

Compensated magnets are of increasing interest for both fundamental research and applications, with their net-zero magnetization leading to ultrafast dynamics and robust order. To understand and control this order, nanoscale mapping of local domain structures is necessary. One of the main routes to mapping antiferromagnetic order is X-ray magnetic linear dichroism (XMLD), which probes the local orientation of the Néel vector. However, XMLD imaging typically suffers from weak contrast and has mainly been limited to surface-sensitive techniques. Here, we harness coherent diffractive imaging to map the complex XMLD spectroscopically,  and identify the phase linear dichroism as a high-contrast, high-resolution mechanism for imaging magnetic order. By applying X-ray spectro-ptychography to a model sample, we retrieve the full complex XMLD spectrum. Combining this with hierarchical clustering, we resolve the spatial distribution of probed domains by their distinct spectral signatures, providing a robust method for analyzing magnetic configurations with weak signals. Our results show that phase contrast is significantly stronger than the corresponding absorption contrast, offering higher spatial-resolution magnetic imaging. This approach establishes a reliable, element- and orbital-sensitive tool for studying compensated magnets.

\end{abstract}

\maketitle

\section{Introduction}

Compensated magnets are systems with spin arrangements resulting in minimal or net-zero magnetization \cite{reichlova2024COMPENSATEDmagnets}. These materials are promising due to their intrinsic properties, including being robust to stray magnetic fields, ultra-fast spin dynamics in the terahertz range, rapid domain wall motion \cite{din2024antiferromagnetic,baltz2018AFMspintronics,jungwirth2016AFspintronics}, and diverse electronic behaviors \cite{chen2024AFMspintronics,siddiqui2020metallicAFM,meer2023AFMoxides,maca2012CuMnAs,ryan2009cAFMsuperconductor}. Once limited to exchange-coupled pinning layers in ferromagnet-based devices, they now play primary roles in antiferromagnetic (AF) spintronics \cite{jungwirth2016AFspintronics,din2024antiferromagnetic,baltz2018AFMspintronics,chen2024AFMspintronics}, driven by breakthroughs in electrical control of the Néel vector orientation \cite{wadley2016electrical,arpaci2021observation,bodnar2019imagingMn2Au}, recent discoveries of antiferromagnetic topological spin structures (e.g., meron-antimeron pairs \cite{amin2023merons}) as well as magnetoelectric effects \cite{kriegner2016AMR,qin2023roomTMR}. 

As AF spintronics advances, applications demand precise control of nanoscale magnetic configurations. Mapping AF domains, identifying pinning centers, and tracking their motion are essential for understanding these systems’ response to electrical currents, especially for homogeneous and reversible domain switching \cite{grzybowski2017AFimagingPEEM,sapozhnik2018dMn2AuPEEM,din2024antiferromagnetic}. These demands for nanoscale characterization techniques grow as interest expands to more complex compensated magnets such as altermagnets, which combine the net-zero magnetization of antiferromagnets with the time-reversal symmetry breaking, and spintronic effects of ferromagnets\cite{vsmejkal2022altermagnets}. 

Despite their potential, the net zero magnetization of compensated magnets poses challenges for conventional techniques that are often sensitive to the magnetization or magnetic induction. X-ray magnetic linear dichroism (XMLD) provides a mechanism to probe local Néel vector orientations \cite{kuiper1993XMLDantiferro,stohr1999,stohr2006magnetism}. So far, most studies have relied on the highly sensitive surface-specific X-ray photoemission electron microscopy (XPEEM) which allows for the nanoscale mapping of the typically weak XMLD contrast \cite{kunevs2003XMLD,kunevs2004XMLDPy,arenholz2006XMLD}. However, imaging beyond the surface requires the development of transmission-based techniques, for which the weak contrast and small size of the domains can be challenging. As a result, suitable methods with high sensitivity and spatial resolution are required.

Here, we present high-contrast imaging of magnetic orientations using coherent XMLD spectro-ptychography. By combining phase imaging with unsupervised machine learning, we identify low-contrast magnetic domains and demonstrate that phase linear dichroic imaging offers higher spatial resolution and signal-to-noise ratio than absorption-based contrast for nanoscale studies of compensated magnets. The high-contrast phase linear dichroism, combined with the potential of diffraction-limited coherent imaging, offers a route to sub-\SI{10}{nm} resolution of ferromagnetic and antiferromagnetic configurations. Moreover, integrating coherent diffractive imaging with advanced computational methods to identify magnetic configurations has promising implications for compensated magnets, key for advancing spintronic applications.  

\subsection{X-ray Magnetic Linear Dichroism (XMLD)}
\label{subsec:XMLD_definition}

When magnetic materials are probed with polarized X-rays in the vicinity of an absorption edge, their scattering and absorption cross-sections depend on the relative orientation of the incident X-ray polarization to the material’s magnetic axis. Known as resonant magnetic scattering (see appendix~\ref{app:res_scattering}), these contrast mechanisms provide a route to study magnetic configurations with element- and atomic orbital-selectivity. There are two main contrast mechanisms: first, X-ray Magnetic Circular Dichroism (XMCD), which probes net magnetic moments along the beam direction - and which is therefore ideal for probing ferromagnets. Second, XMLD is sensitive to the orientation of moments in the plane perpendicular to the beam, making it well-suited for studying both ferromagnetic and antiferromagnetic systems with net-zero magnetization.

Combining XMLD with nanoscale microscopy has revealed key features of compensated magnets, including the response of domains to external stimuli \cite{sapozhnik2018afmPEEM,lee2024}, as well as the structure of domain walls \cite{Arai2012} and topological structures~ \cite{amin2023merons}. The vast majority of XMLD microscopy has made use of highly sensitive X-PEEM, that offers surface imaging typically with \(\sim\) 20-\SI{30}{nm} resolution. While techniques like Scanning Transmission X-ray Microscopy (STXM)~ \cite{luo2023} and Fourier Transform Holography~ \cite{Harrison2024} probe bulk configurations in transmission geometry, XPEEM remains dominant due to its high sensitivity, ideal for the measurement of the typically weak XMLD contrast. With growing interest in topological textures and emergent phenomena in new materials like altermagnets~ \cite{vsmejkal2022altermagnets}, there is growing demand for imaging methods with high contrast, spatial resolution, and bulk sensitivity, that can be performed under complex \textit{in situ} conditions. To this end, the development of coherent imaging techniques like X-ray ptychography, which offer prospects of diffraction-limited resolution and high sensitivity, represents promising opportunities.  

\begin{figure}[h]
    \centering
    \includegraphics[width=\linewidth]{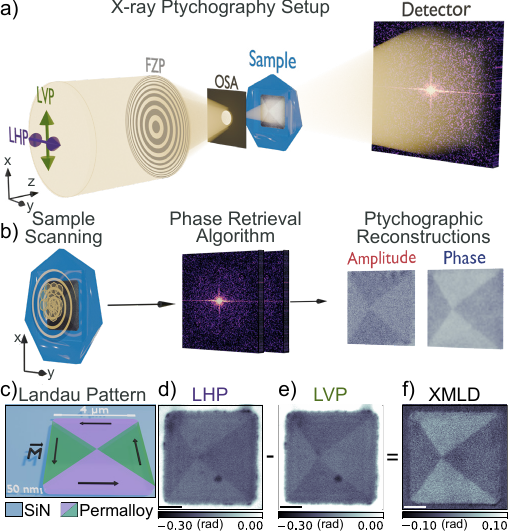}
    \caption{Dichroic X-ray ptychography setup and XMLD imaging. (a) Schematic of the X-ray ptychography setup: A coherent X-ray beam with LVP or LHP polarization is focused through a Fresnel Zone Plate (FZP) and Order Sorting Aperture (OSA) onto the sample, with diffraction patterns captured by a far-field positioned detector. (b) The permalloy thin film on a \(\text{Si}_3\text{N}_4\) membrane (c) forms a Landau pattern and is scanned in overlapping positions along a spiral path. Each diffraction pattern is processed with an iterative phase retrieval algorithm to yield images with phase and amplitude contrast. (d, e) Subtraction of LHP and LVP images yields the XMLD contrast image (f), highlighting magnetic domains. Scale bar: \SI{1}{\um}.}
    \label{fig:setup}
\end{figure}

\subsection{X-ray Ptychography}

X-ray ptychography is a scanning form of coherent diffractive imaging (CDI) that can overcome key limitations of traditional X-ray microscopy approaches, for example going beyond the limits of X-rays optics to diffraction-limited spatial resolutions. Employing robust phase-retrieval algorithms, X-ray ptychography reconstructs images from far-field diffraction patterns, allowing full recovery of the complex transmission function of an object as $T(E) = A(E)\,e^{i\phi(E)} $, enabling both amplitude ($A$) and phase ($\phi$) contrast imaging. Notably, as a lensless imaging technique, the resolution is not limited by X-ray optics, thus pushing closer to diffraction-limited resolutions~ \cite{shapiro2014,Sun2021}.

In recent years, X-ray ptychography has proven to be highly advantageous for magnetic materials, allowing for the imaging of multiferroic spin spirals with high spatial resolution~ \cite{butcher2024MultferroicPtycho}, providing high sensitivity for nanoscale imaging of large systems with hard X-rays~ \cite{donnelly2016hardXRAY}, which paved the way to vector magnetic tomography~ \cite{donnelly2017vectorTOMO}, and providing single polarization dichroic imaging with multimodal reconstruction methods \cite{DiPietro_singlepolPtycho}. By providing access to the complex XMCD signal, the benefits of phase dichroism have become clear, allowing for the study of thicker samples that were previously challenging for transmission methods~ \cite{Jeffrey2024}. While highly effective for magnetic circular dichroism, the application of coherent imaging to magnetic linear dichroism remains relatively unexplored.

 \begin{figure*}[t!]
    \centering
    \includegraphics[width=\linewidth]{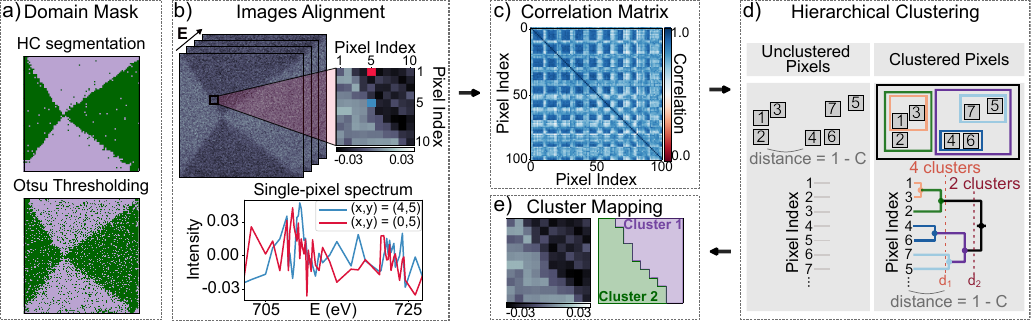}
    \caption{ Magnetic domain segmentation via hierarchical clustering. (a) Mask obtained with hierarchical clustering (top), distinguishing two clusters corresponding to regions with perpendicular magnetization. For comparison, a mask from Otsu Thresholding (bottom) shows higher noise. (b-e) Steps for determining magnetic domain distribution using a correlation-based hierarchical clustering algorithm~ \cite{LUCASphdthesis}. (b) Ptychography images at multiple energies are aligned, allowing pixel-wise spectra to be obtained. The inset shows a zoomed region with $10 \times 10$ pixels in which two pixels are highlighted (in blue and red). The single-pixel spectra from the highlighted pixels are then presented. (c) The calculated pixel-to-pixel correlation matrix ($100 \times 100$) of this region provides the correlation values ($C$) between each single-pixel spectra that will be used for the clustering. (d) Schematic of the hierarchical clustering process, where pixels are incrementally clustered based on the distance metric defined as $1 - C$. (e) This clustering yields a segmentation mask for the zoomed region, mapping the two magnetic domains. }
    \label{fig:HC}
\end{figure*}

\section{XMLD Spectro-ptychography}
\label{subsec:exp_setup}

To determine whether there may be benefits of coherent diffractive imaging for magnetic linear dichroic imaging, we map the full complex XMLD signal by performing spectro-ptychography on a multi-domain system. As detailed in Appendix\ref{app:exp_setup}, experiments were performed at the I08-1 instrument of Diamond Light Source, using a coherent X-ray beam with Linear Vertical (LVP) or Linear Horizontal Polarization (LHP). A \SI{1}{\um} probe was defined at the sample position using a Fresnel Zone Plate and Order Sorting Aperture (Fig.\ref{fig:setup}a). Coherent diffraction patterns were collected in the far field on a sCMOS detector by scanning the sample in overlapping spiral paths (Fig.\ref{fig:setup}a,b) across the Fe $L_{2,3}$ edges. Images were reconstructed using the python-based framework PtyPy~ \cite{ptypy}.

XMLD spectro-ptychography was performed on a lithographically patterned permalloy (\(\text{Ni}_{80}\text{Fe}_{20}\)) micro-square (Appendix~\ref{app:sample}) exhibiting a Landau pattern: a vortex configuration with triangular in-plane ferromagnetic domains induced by shape anisotropy (Fig.~\ref{fig:setup}c). This configuration provides an ideal test sample for the experiment due to its well-characterized in-plane magnetic domain regions with perpendicular magnetization. While this is not an antiferromagnetic sample, the results translate to such systems, as the XMLD is probing the spin-axis orientation and not the magnetization direction. We obtained single-polarization images (Fig.~\ref{fig:setup}d,e), where LVP and LHP data combine electronic density and magnetic contrast. To isolate the magnetic contribution, we align, normalize, and subtract the two images to calculate the XMLD image (Fig.~\ref{fig:setup}b), as explained in Appendix~\ref{app:xmld_images_calc}.

The XMLD image (Fig.\ref{fig:setup}f) reveals triangular domains with opposite contrast, corresponding to horizontal (bright) and vertical (dark) in-plane magnetization (Fig.\ref{fig:setup}c), consistent with the expected Landau pattern. The contrast arises from linearly polarized light interacting strongly with magnetization parallel to the polarization direction and weakly with perpendicular magnetization. At the center, the triangular domains meet at a vortex core, slightly shifted due to the presence of a small magnetic field in the setup. We next analyze the energy dependence of the magnetic contrast comparing phase and absorption spectra.

\subsection{Spectroscopic Hierarchical Clustering}
\label{subsec:HC}

In order to extract the energy dependence of the complex dichroism, the identification of domains in the sample is necessary. While for strong signals, domain segmentation is feasible through image thresholding~ \cite{Jeffrey2024} or by leveraging prior knowledge of the sample’s properties, the application to unknown samples, small signals, and/or fewer statistics can be challenging. Here, although the magnetic domains are visually discernible at the Fe \(L_3\)-edge absorption maximum (\SI{710}{eV}, Fig.\ref{fig:setup}f), the low contrast and high noise leads to high variability in pixel spectra (Fig.\ref{fig:HC}b), complicating standard segmentation techniques such as Otsu Thresholding (lower image in Fig.~\ref{fig:HC}a) and requiring a more sophisticated approach.

To segment the magnetic domains, we apply the unsupervised machine learning technique of hierarchical clustering, bypassing the need for prior knowledge of the sample and achieving robustness to noise through data correlation. Our approach segregates the regions based on a physical property: the absorption/phase X-ray spectrum of each pixel. As illustrated in Figure~\ref{fig:HC}b-e, firstly all normalized single polarization images are aligned via sub-pixel registration methods, ensuring consistent pixel correspondence across different energies (Fig.~\ref{fig:HC}b). Next, each single-pixel spectrum forms a vector, and the vector array is then used to construct a pixel-to-pixel correlation matrix, capturing the degree of similarity (values close to one) or dissimilarity (values approaching zero) between the spectra of individual pixels. In Fig.~\ref{fig:HC}c, the correlation matrix for the $10 \times 10$ pixel region highlighted in the inset of Fig.~\ref{fig:HC}b is shown. The main diagonal values are one (self-correlation), while square-like regions of high correlation represent pixels within the same domain, separated by low-correlation elements indicating different domains.

From this correlation matrix, we define a “distance” metric, calculated as \(1 - C\), where \(C\) is the correlation value. This value quantifies how each pixel's spectral behavior differs from others, with higher distances corresponding to pairs of pixels with low correlation. Using the bottom-up hierarchical clustering approach through the linkage method available in SciPy~ \cite{2020SciPy}, we group the most similar pixels based on the correlation-based metric, progressively clustering pixels with smaller distances (larger correlation)~ \cite{LUCASphdthesis} as displayed in Fig.~\ref{fig:HC}d. Plotting the two clusters with the largest distance—indicating low inter-cluster correlation but strong intra-cluster correlation—reveals the perpendicularly oriented magnetic domains of the Landau pattern, as shown in the zoomed region of Fig.~\ref{fig:HC}e.

\begin{figure*}[ht]
    \centering
    \includegraphics[width=0.8\linewidth]{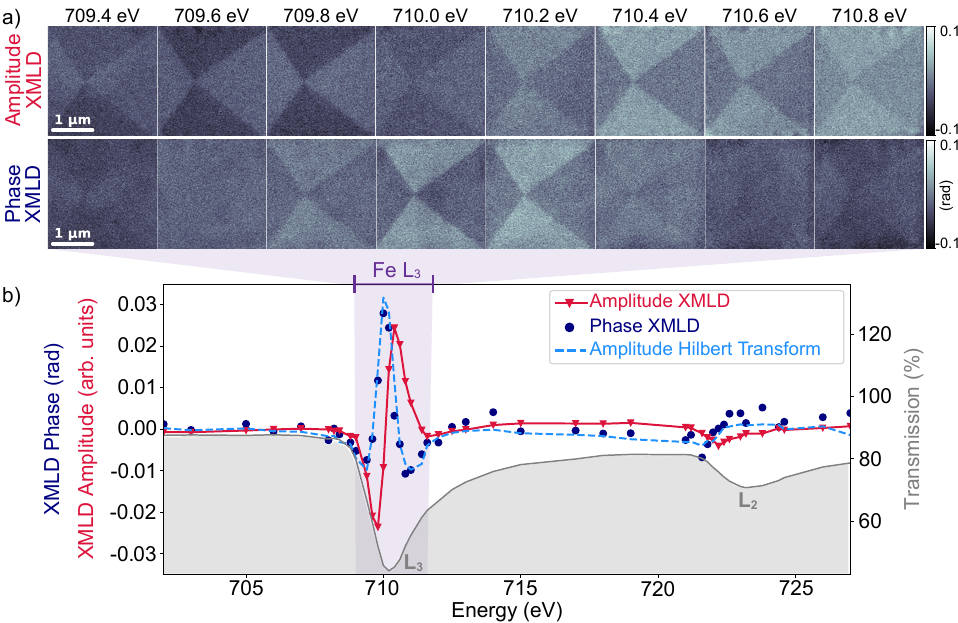}
    \caption{Energy-dependent XMLD ptychography imaging and complex spectrum of the Landau pattern. (a) XMLD ptychography images at the Fe \( L_3 \)-edge (\( 709.4 \) to \( 710.8 \) eV) showing amplitude (top) and phase (bottom) contrast, highlighting the magnetic domain structure in the Landau state. (b) Spectra at the Fe \( L_3 \) and \( L_2 \) edges, with the amplitude transmission spectrum (grey), phase XMLD spectrum (blue circles), amplitude XMLD spectrum (red triangles), and the Hilbert transform of the amplitude (blue dashed line), yielding the phase signal.}
    \label{fig:spec}
\end{figure*}

Applying this spectroscopic hierarchical clustering segmentation to the entire sample provides a spatial distribution of the magnetic domains, with the domains and their boundaries identified with high fidelity (Fig.~\ref{fig:HC}a). Compared to Otsu thresholding (Fig.\ref{fig:HC}a lower image), hierarchical clustering results in significantly improved segmentation. Notably, hierarchical clustering method segments based solely on spectral similarity without the need for prior information of the sample, thus resulting in high-resolution mapping of domains within the sample. We next use this segmentation to extract quantitative, cluster (i.e. orientation)-specific XMLD spectra as detailed in Appendix~\ref{subsec:XMLD_calc}.

\subsection{Complex XMLD Spectra}
\label{subsec:XMLD_spec_ptycho}

After identifying the domain-specific spectra with hierarchical clustering (App.~\ref{subsec:XMLD_calc}), we analyze the full complex XMLD spectrum. In Fig.~\ref{fig:spec}a we plot XMLD ptychography images at eight energy points across the Fe \(L_3\) edge, along with the phase (blue circles) and amplitude (red triangles) XMLD spectra at the Fe \(L_3\) and \(L_2\) absorption edges in Fig.~\ref{fig:spec}b, together with with the sample’s transmission spectrum (gray). The amplitude and phase spectra are intrinsically related through the Kramers-Kronig relations (mathematically equivalent to a Hilbert transform). The phase XMLD spectrum is compared to the Hilbert transform of the amplitude signal (blue dashed line) in Fig.~\ref{fig:spec}b, where a good agreement can be seen, highlighting the quantitative nature of phase dichroic ptychography for spectroscopy.

The amplitude spectrum (Fig.~\ref{fig:spec}b) features the expected~ \cite{kunevs2004XMLDPy,schwickert1998FeEdgeXMLD} prominent signal inversion across the \(L_3\) edge, with a dip at \(\SI{709.8}{\eV}\) and a peak at \(\SI{710.4}{\eV}\), while the signal approaches zero between the \(L_3\) and \(L_2\) edges. Conversely, the phase spectrum has a single peak around \(710\) to \(\SI{710.2}{\eV}\), where the sharpest magnetic contrast appears, with two subtle dips on either side of the main peak, visible as small contrast inversions in the images at \(\SI{709.4}{\eV}\) and \(\SI{710.8}{\eV}\).

Notably, the energy dependence of the XMLD signal contrasts with that of XMCD spectra obtained via spectro-ptychography. In XMCD, the phase spectrum exhibits signal inversion, while the amplitude spectrum shows a narrow peak across the \(L_3\) edge~ \cite{Scherz07,donnelly2016hardXRAY,Jeffrey2024}, with the broader magnetic contrast range of phase XMCD enabling imaging of thicker systems through pre-edge phase dichroism~ \cite{Jeffrey2024}. In contrast, for XMLD, the amplitude presents contrast maxima not aligned with the highest absorption region of the spectrum (\(\sim \SI{710}{eV}\)), potentially allowing the measurement of magnetic contrast in the pre-edge energies associated with higher-transmission. Finally, as seen in the XMLD spectra and imaging (Fig.~\ref{fig:spec}a), phase imaging appears to provide sharper features and higher contrast than amplitude imaging, offering a potential route to higher spatial resolution imaging of antiferromagnetic features.

\subsection{Resolution and Contrast Analysis}
\label{subsec:resolution}

Having identified the potential for phase XMLD imaging, we assess the spatial resolution of phase and amplitude XMLD at their respective maximum contrast energies, \( \SI{710}{eV} \) and \( \SI{710.4}{eV} \), by analyzing intensity profiles across domain boundaries (see Appendix~\ref{app:SNR}). The average measured domain wall width obtained was \( \SI{62 \pm 5}{nm} \) for phase and \( \SI{79 \pm 12}{nm} \) for amplitude, with phase profiles exhibiting sharper boundaries, thus implying a higher spatial resolution. Indeed, Fourier ring correlation (FRC) analysis allowed for estimates of the spatial resolution of \( \SI{86}{nm} \) for the phase and \( \SI{92}{nm} \) for the amplitude. We note that the lower values of resolution provided by the FRC is likely to be due to the absence of high-frequency features in the relatively featureless images,  thus resulting in an underestimation of the spatial resolution. However, using both methods, the spatial resolution of the XMLD phase consistently outperformed that of the amplitude signal. 

We determine the reason for this difference in spatial resolution by calculating the signal-to-noise ratio (SNR) of the phase and amplitude images (App.~\ref{app:SNR}). Here, the phase images consistently exhibit higher SNR, up to 43\% higher than amplitude at the \( L_3 \)-edge. Consequently, to match the SNR of a phase image, twice as many amplitude images would be needed, meaning that phase XMLD represents an efficient method for the high-resolution imaging of ferromagnets and antiferromagnets. Finally, we note that the spatial resolutions in this demonstration do not compete with the highest spatial resolutions demonstrated so far. The spatial resolution in ptychography depends on various factors, including the fraction of incoherent flux, the sample absorption cross-section, the detector dynamic range, and exposure time. When the experiment is optimized for high resolution, ptychography can provide resolutions far exceeding those reported here. In those cases, the higher SNR and spatial resolution of the phase could allow for significant advances in the spatial resolution for imaging antiferromagnetic domain configurations.

\section{Conclusion and Outlooks}

Here we have established X-ray magnetic linear dichroic spectro-ptychography, harnessing linear dichroism and coherent diffractive imaging to achieve high-quality imaging of in-plane ferromagnetic domains. With hierarchical clustering, we perform a physics-informed segmentation of the domains with no prior knowledge of the sample and extract high-quality complex XMLD spectra, recovering both the known amplitude spectrum, and mapping the phase XMLD spectrum. Remarkably, we find that phase XMLD imaging provides sharper features, higher spatial resolution, and up to \(43\%\) higher signal-to-noise ratio compared to amplitude imaging. 

These findings establish XMLD spectro-ptychography as a powerful and sensitive tool for investigating a wide range of systems. Although demonstrated on a ferromagnet, this technique has broad applicability to a variety of materials, including antiferromagnets, and the emerging class of altermagnets~ \cite{amin2024altermagnetism}, which promise to combine the robustness of antiferromagnets with the strong spintronic effects of ferromagnets. The use of hierarchical clustering in this method enables detailed analysis of nanoscale spectral features, which could be applied to the spectroscopic analysis of more intricate magnetic textures. Indeed, the principle of hierarchical clustering could be applied to spectroscopic data in general, aiding in the identification of different phases within materials, ranging from oxidation states in energy materials to ferroic order in multiferroics. As a photon-in photon-out technique, it is particularly well-suited for \textit{in situ} measurements, including experiments under applied magnetic fields and currents, such as tracking domain evolution under external stimuli. Moreover, as a transmission-based technique, it opens the door to higher dimensional imaging such as tomography. For instance, the power of the phase XMLD ptychography could be combined with linear dichroic orientation tomography~ \cite{Apseros2024XMLDtomo}, to provide high spatial resolution tomography of antiferromagnets, mapping three-dimensional antiferromagnetic configurations in space. 

\section{Acknowledgments}

This work was carried out with the support of Diamond Light Source, instrument \(\text{I}08\text{-}1\) (proposal \(\text{MG}30417\)). M.R.F acknowledges financial support from both the Max Planck Society under the Max Planck Partner Group R. D. dos Reis of the MPI for Chemical Physics of Solids, Dresden, Germany, and the Serrapilheira Institute (Grant Number G-1709-17301). C.D., J.N., and S.R.G acknowledge funding from the Max Planck Society Lise Meitner Excellence Program and funding from the European Research Council (ERC) under the ERC Starting Grant No. 3DNANOQUANT 101116043. A. A. acknowledges financial support from the Swiss National Science Foundation (SNSF), project number \(200021\_192162\). We acknowledge L. H. Francisco for the fruitful discussions on hierarchical clustering. 

\appendix
\section*{Appendices}
\addcontentsline{toc}{section}{Appendices}
\renewcommand{\thesubsection}{\Alph{subsection}}

\subsection{X-ray Resonant Magnetic Scattering}
\label{app:res_scattering}

The X-ray scattering and absorption cross-sections of magnetic materials depends on the relative orientation of the magnetic axis to the incident polarization. This is described by the resonant magnetic scattering factor \( f(E, \bm{r}) \) for dipole-allowed transitions \cite{lovesey1996x,van2008soft}:
\begin{equation}
    \label{eq:mag_scat_fac}
    \begin{split}
          f(E,\bm{r})\, = & \,\underbrace{f_c(E)(\bm{\epsilon}_f^*\cdot \bm{\epsilon}_i)}_{\text{Charge}}\,-\underbrace{if_m^{(1)}(E)(\bm{\epsilon}_f^*\times \bm{\epsilon}_i)\cdot \bm{m}(\bm{r})}_{\text{Circular Dichroism}} \, \\
        & + \underbrace{f^{(2)}_m(E) \big(\bm{\epsilon}_f^*\cdot \bm{m} ( \bm{r}) \big)\big(\bm{\epsilon}_i\cdot \bm{m}(\bm{r})\big)}_{\text{Linear Dichroism}}
    \end{split}
\end{equation}
Here, $\bm{\epsilon_i}$ and $\bm{\epsilon_f}$ are the incident and outgoing polarization vectors, $\bm{m(r)}$ is the magnetization vector, and $f_c$, $f_m^{(1)}$, and $f_m^{(2)}$ are the scattering factors. From equation~(\ref{eq:mag_scat_fac}), XMCD is sensitive to net magnetization (\(\bm{m} ( \bm{r})\)-dependence) along the beam path direction. In contrast, XMLD probes the spin-axis orientation (\( \langle m^2 \rangle \text{-dependence}\)) perpendicular to the beam propagation direction, making it sensitive to the magnetic structure of both antiferromagnetic and ferromagnetic systems.

\subsection{X-ray Ptychography Experimental Setup and Images Reconstruction}
\label{app:exp_setup}

The X-ray ptychography experiments were conducted at the I08-1 instrument of Diamond Light Source. A Fresnel Zone Plate (FZP) with a diameter of \SI{333}{\micro\meter} and an outermost zone width of \SI{70}{\nano\meter} followed by an Order Sorting Aperture (OSA) of \SI{25}{\micro\meter} focused the beam to a spot size of approximately \SI{1}{\micro\meter} at the sample position. Using piezoelectric scanners, the sample was scanned in a spiral pattern. At each position, diffraction patterns were acquired with a \SI{40}{\milli\second} exposure time, utilizing a sCMOS detector (model EUV-Enhanced GSENSE400 BSI, Axis Photonique Inc.) forming a \(2048 \times 2048\) pixel array with \SI{11}{\micro\meter} pixel size. The detector was positioned \SI{72}{\milli\meter} downstream from the sample. The images were then cropped to \(1024 \times 1024\) pixels, and binned to \(512 \times 512\) pixels, resulting in an effective pixel size of \SI{22}{\micro\meter}. Measurements were taken at 52 energy points across the Fe \(L_3\) and \(L_2\) absorption edges, with both linear horizontal and vertical polarizations. Image reconstruction was performed using the python-based framework PtyPy~ \cite{ptypy}, combining Relaxed Average Alternating Reflection (RAAR) and Maximum Likelihood (ML) engines, running 2000 iterations each, with a fixed pixel size of \SI{22.611}{\nano\meter} for all energies.

\subsection{Sample}
\label{app:sample}
The \SI{50}{nm}-thick permalloy ($\text{Ni}_{80}\text{Fe}_{20}$) film was fabricated by e-beam lithography, DC magnetron sputtering and lift off on a \(\text{Si}_3\text{N}_4\) transparent membrane. The patterned structure was a square with lateral dimensions of \SI{4}{\mu m}, responsible for inducing the desired shape anisotropy displayed in Fig.~\ref{fig:setup}a.
\subsection{Ptychography Images Normalization}
\label{app:xmld_images_calc}
Before calculating XMLD, all single-polarization images (LHP or LVP) were aligned using sub-pixel alignment methods~ \cite{scikit-image}. Phase-contrast images were corrected for phase ramping. All images were normalized by the background intensity according to equations~(\ref{eq:A_norm}) and~(\ref{eq:Phi_norm}).
\begin{equation}
\label{eq:A_norm}
    A_{\text{NORM}} = \frac{A_{\text{sample}}}{A_{\text{background}}}
\end{equation}
\begin{equation}
\label{eq:Phi_norm}
    \phi_{\text{NORM}} = \phi_{\text{sample}} - \phi_{\text{background}}
\end{equation}
The amplitude and phase XMLD images were then computed by taking the difference between the LHP and LVP images, as outlined in equations~(\ref{eq:xmld_amp}) and~(\ref{eq:xmld_ph}).
\begin{equation}
    \begin{split}
    \label{eq:xmld_amp}
        A_{\text{XMLD}} & = \text{ln}\left(-A^{\text{LHP}}_{\text{NORM}}\right) - \text{ln}\left(-A^{\text{LVP}}_{\text{NORM}}\right) \\
        & \propto \Im[f^{(2)}_m]({m_x}^{2} - {m_y}^{2})
    \end{split}
\end{equation}
\begin{equation}
\label{eq:xmld_ph}
    \begin{split}
        \phi_{\text{XMLD}} &= \phi^{\text{LHP}}_{\text{NORM}} - \phi^{\text{LVP}}_{\text{NORM}} \\
        & \propto \Re [f^{(2)}_m]({m_x}^{2} - {m_y}^{2})
    \end{split}
\end{equation}
where \(\Re [f^{(2)}_m]\) and \(\Im [f^{(2)}_m]\) correspond to the real and imaginary parts of the complex magnetic scattering factor in equation~(\ref{eq:mag_scat_fac}).

\subsection{XMLD Spectra Calculation}
\label{subsec:XMLD_calc}
\begin{figure}[h!]
    \centering
    \includegraphics[width=0.8\linewidth]{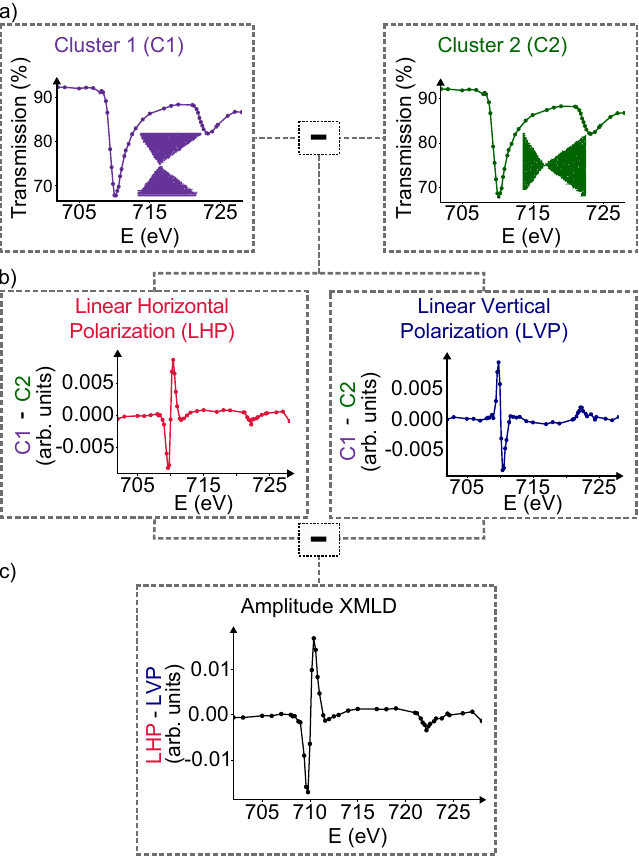}
    \caption{ Steps from the domain mask to XMLD spectrum: (a) Using the domain mask obtained with hierarchical clustering, the transmission spectra are obtained separately for each domain (cluster) and polarization. (b) The difference between these spectra displays the absorption contrast between the magnetic domain regions; this is done separately for Linear Horizontal (LHP) and Linear Vertical Polarization (LVP). (c) We display the XMLD amplitude spectrum defined as the difference between LHP and LVP spectra.}
    \label{fig:spec_calc}
\end{figure}

\begin{figure}[h!]
    \centering
    \includegraphics[width=0.8\linewidth]{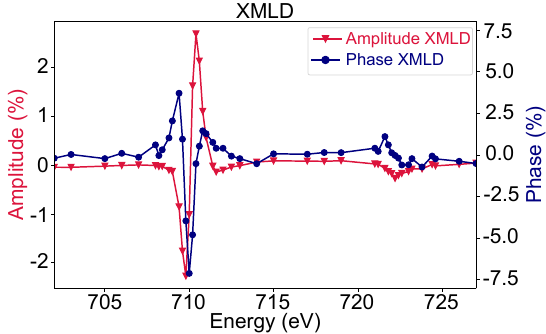}
    \caption{Complex X-ray Magnetic Linear Dichroism (XMLD) spectra for phase contrast (blue circles) and amplitude contrast (red triangles), expressed as percentages according to Equation~(\ref{eq:xmld_perc}).}
    \label{fig:XMLD_perc}
\end{figure}

Using the segmentation mask from hierarchical clustering (Figure~\ref{fig:HC}a), we extract domain-specific spectra by analyzing each cluster. The steps for obtaining the complex XMLD spectra are outlined in Figures~\ref{fig:spec_calc}a to c. First, we calculate the X-ray transmission spectra by integrating intensities across each domain and polarization separately (Figure~\ref{fig:spec_calc}a). Next, we compute the difference between the transmission spectra of each domain, yielding the single-polarization spectra shown in Figure~\ref{fig:spec_calc}b, which highlights the absorption contrast between regions with perpendicular magnetization. Finally, the XMLD spectrum is defined as the difference between the LHP and LVP spectra, producing a robust signal with a characteristic sharp change at the \(L_3\)-edge, as seen in Figure~\ref{fig:spec_calc}c. The spectra in Figures~\ref{fig:spec_calc} and~\ref{fig:spec} are shown in arbitrary units but can be represented as percentages, as described in Equation~(\ref{eq:xmld_perc}), yielding the curves in Figure~\ref{fig:XMLD_perc}. This demonstrates that phase contrast achieves a higher XMLD signal percentage (\(\sim 7.1 \%\)) compared to amplitude contrast (\(\sim 2.7 \%\)).
\begin{equation}
    \label{eq:xmld_perc}
    I_{\text{XMLD}}(\%) = \left ( \frac{I_{\text{LHP}}-I_{\text{LVP}}}{I_{\text{LHP}}+I_{\text{LVP}}} \right) \times 100 
\end{equation}
\subsection{Resolution and Signal-to-Noise Ratio Calculation}
\label{app:SNR}
\begin{figure}
   \centering
    \includegraphics[width=0.8\linewidth]{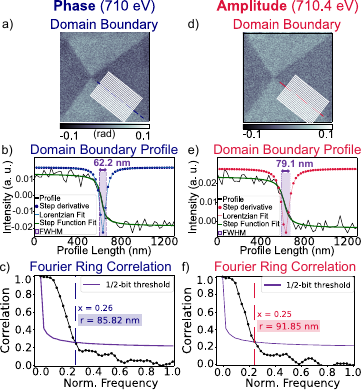}
    \caption{Spatial resolution analysis at peak contrast energies for phase imaging (a–c) at \( \SI{710}{eV} \) and amplitude imaging (d–f) at \( \SI{710.4}{eV} \). (a, d) Line profile distributions perpendicular to domain boundaries. (b, e) Average intensity profiles (black) with arctangent-function fits (dark green) and their derivatives (circles); domain boundary widths are defined as the FWHM of the derivative (\(\SI{62.2}{nm}\) for phase and \(\SI{79.1}{nm}\) for amplitude). (c, f) Fourier ring correlation with a \(1/2\)-bit threshold, yielding resolutions of \(\SI{85.8}{nm}\) for phase and \(\SI{91.9}{nm}\) for amplitude.}
    \label{fig:res} 
\end{figure}

To assess the spatial resolution, we analyze intensity profiles across four domain boundaries. Given high noise levels, we average multiple profiles taken perpendicular to these boundaries (Figs.~\ref{fig:res}a,d). Each profile is fitted with an arctangent function, and the boundary width is calculated as the full width at half maximum (FWHM) from the profile fit derivative, yielding an average domain wall width of \( \SI{62 \pm 5}{nm} \) for phase and \( \SI{79 \pm 12}{nm} \) for amplitude. Fourier ring correlation (FRC) analysis was performed to images using the \(1/2\)-bit threshold criterion~ \cite{VANHEEL2005FRC}, resulting in resolutions of \(\SI{85}{\nano\meter}\) for phase and \(\SI{91}{\nano\meter}\) for amplitude images.

\begin{figure}[h]
    \centering
    \includegraphics[width=0.9\linewidth]{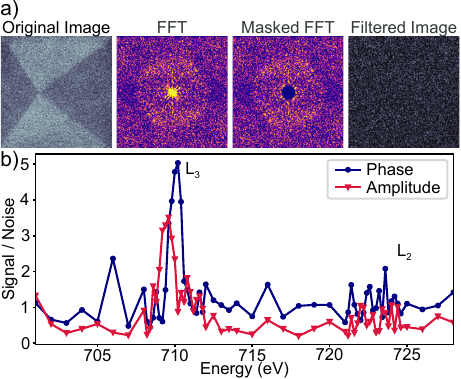}
    \caption{(a) High-frequency noise extraction process for SNR calculation: The FFT of the XMLD image is masked to remove low-frequency components. The inverse FFT yields the filtered image for standard deviation computation. (b) SNR as a function of energy around the Fe $L_{2,3}$-edges, for amplitude (red triangles) and phase (blue circles).}
    \label{fig:SNR}
\end{figure}

We calculate the signal-to-noise ratio (SNR) of the XMLD images as the ratio of averaged signal to the standard deviation of high-frequency noise. High-frequency noise is extracted by performing a fast Fourier transform (FFT) of the XMLD image, masking the low-frequency signal associated with magnetic domains, and applying an inverse FFT to obtain the filtered image. The SNR, shown in Fig.~\ref{fig:SNR}b, is plotted for amplitude (red triangles) and phase (blue circles) as a function of energy. The phase SNR peaks at approximately 5.0 at \SI{710.2}{eV}, while the maximum amplitude SNR of 3.5 occurs at \SI{709.6}{eV}.

\begin{thebibliography}{47}%
\makeatletter
\providecommand \@ifxundefined [1]{%
 \@ifx{#1\undefined}
}%
\providecommand \@ifnum [1]{%
 \ifnum #1\expandafter \@firstoftwo
 \else \expandafter \@secondoftwo
 \fi
}%
\providecommand \@ifx [1]{%
 \ifx #1\expandafter \@firstoftwo
 \else \expandafter \@secondoftwo
 \fi
}%
\providecommand \natexlab [1]{#1}%
\providecommand \enquote  [1]{``#1''}%
\providecommand \bibnamefont  [1]{#1}%
\providecommand \bibfnamefont [1]{#1}%
\providecommand \citenamefont [1]{#1}%
\providecommand \href@noop [0]{\@secondoftwo}%
\providecommand \href [0]{\begingroup \@sanitize@url \@href}%
\providecommand \@href[1]{\@@startlink{#1}\@@href}%
\providecommand \@@href[1]{\endgroup#1\@@endlink}%
\providecommand \@sanitize@url [0]{\catcode `\\12\catcode `\$12\catcode `\&12\catcode `\#12\catcode `\^12\catcode `\_12\catcode `\%12\relax}%
\providecommand \@@startlink[1]{}%
\providecommand \@@endlink[0]{}%
\providecommand \url  [0]{\begingroup\@sanitize@url \@url }%
\providecommand \@url [1]{\endgroup\@href {#1}{\urlprefix }}%
\providecommand \urlprefix  [0]{URL }%
\providecommand \Eprint [0]{\href }%
\providecommand \doibase [0]{https://doi.org/}%
\providecommand \selectlanguage [0]{\@gobble}%
\providecommand \bibinfo  [0]{\@secondoftwo}%
\providecommand \bibfield  [0]{\@secondoftwo}%
\providecommand \translation [1]{[#1]}%
\providecommand \BibitemOpen [0]{}%
\providecommand \bibitemStop [0]{}%
\providecommand \bibitemNoStop [0]{.\EOS\space}%
\providecommand \EOS [0]{\spacefactor3000\relax}%
\providecommand \BibitemShut  [1]{\csname bibitem#1\endcsname}%
\let\auto@bib@innerbib\@empty
\bibitem [{\citenamefont {Reichlova}\ \emph {et~al.}(2024)\citenamefont {Reichlova}, \citenamefont {Kriegner}, \citenamefont {Mook}, \citenamefont {Althammer},\ and\ \citenamefont {Thomas}}]{reichlova2024COMPENSATEDmagnets}%
  \BibitemOpen
  \bibfield  {author} {\bibinfo {author} {\bibfnamefont {H.}~\bibnamefont {Reichlova}}, \bibinfo {author} {\bibfnamefont {D.}~\bibnamefont {Kriegner}}, \bibinfo {author} {\bibfnamefont {A.}~\bibnamefont {Mook}}, \bibinfo {author} {\bibfnamefont {M.}~\bibnamefont {Althammer}},\ and\ \bibinfo {author} {\bibfnamefont {A.}~\bibnamefont {Thomas}},\ }\bibfield  {title} {\bibinfo {title} {Role of topology in compensated magnetic systems},\ }\href@noop {} {\bibfield  {journal} {\bibinfo  {journal} {APL Materials}\ }\textbf {\bibinfo {volume} {12}} (\bibinfo {year} {2024})}\BibitemShut {NoStop}%
\bibitem [{\citenamefont {Dal~Din}\ \emph {et~al.}(2024)\citenamefont {Dal~Din}, \citenamefont {Amin}, \citenamefont {Wadley},\ and\ \citenamefont {Edmonds}}]{din2024antiferromagnetic}%
  \BibitemOpen
  \bibfield  {author} {\bibinfo {author} {\bibfnamefont {A.}~\bibnamefont {Dal~Din}}, \bibinfo {author} {\bibfnamefont {O.}~\bibnamefont {Amin}}, \bibinfo {author} {\bibfnamefont {P.}~\bibnamefont {Wadley}},\ and\ \bibinfo {author} {\bibfnamefont {K.}~\bibnamefont {Edmonds}},\ }\bibfield  {title} {\bibinfo {title} {Antiferromagnetic spintronics and beyond},\ }\href@noop {} {\bibfield  {journal} {\bibinfo  {journal} {npj Spintronics}\ }\textbf {\bibinfo {volume} {2}},\ \bibinfo {pages} {25} (\bibinfo {year} {2024})}\BibitemShut {NoStop}%
\bibitem [{\citenamefont {Baltz}\ \emph {et~al.}(2018)\citenamefont {Baltz}, \citenamefont {Manchon}, \citenamefont {Tsoi}, \citenamefont {Moriyama}, \citenamefont {Ono},\ and\ \citenamefont {Tserkovnyak}}]{baltz2018AFMspintronics}%
  \BibitemOpen
  \bibfield  {author} {\bibinfo {author} {\bibfnamefont {V.}~\bibnamefont {Baltz}}, \bibinfo {author} {\bibfnamefont {A.}~\bibnamefont {Manchon}}, \bibinfo {author} {\bibfnamefont {M.}~\bibnamefont {Tsoi}}, \bibinfo {author} {\bibfnamefont {T.}~\bibnamefont {Moriyama}}, \bibinfo {author} {\bibfnamefont {T.}~\bibnamefont {Ono}},\ and\ \bibinfo {author} {\bibfnamefont {Y.}~\bibnamefont {Tserkovnyak}},\ }\bibfield  {title} {\bibinfo {title} {Antiferromagnetic spintronics},\ }\href@noop {} {\bibfield  {journal} {\bibinfo  {journal} {Reviews of Modern Physics}\ }\textbf {\bibinfo {volume} {90}},\ \bibinfo {pages} {015005} (\bibinfo {year} {2018})}\BibitemShut {NoStop}%
\bibitem [{\citenamefont {Jungwirth}\ \emph {et~al.}(2016)\citenamefont {Jungwirth}, \citenamefont {Marti}, \citenamefont {Wadley},\ and\ \citenamefont {Wunderlich}}]{jungwirth2016AFspintronics}%
  \BibitemOpen
  \bibfield  {author} {\bibinfo {author} {\bibfnamefont {T.}~\bibnamefont {Jungwirth}}, \bibinfo {author} {\bibfnamefont {X.}~\bibnamefont {Marti}}, \bibinfo {author} {\bibfnamefont {P.}~\bibnamefont {Wadley}},\ and\ \bibinfo {author} {\bibfnamefont {J.}~\bibnamefont {Wunderlich}},\ }\bibfield  {title} {\bibinfo {title} {Antiferromagnetic spintronics},\ }\href@noop {} {\bibfield  {journal} {\bibinfo  {journal} {Nature nanotechnology}\ }\textbf {\bibinfo {volume} {11}},\ \bibinfo {pages} {231} (\bibinfo {year} {2016})}\BibitemShut {NoStop}%
\bibitem [{\citenamefont {Chen}\ \emph {et~al.}(2024)\citenamefont {Chen}, \citenamefont {Liu}, \citenamefont {Zhou}, \citenamefont {Meng}, \citenamefont {Wang}, \citenamefont {Duan}, \citenamefont {Zhao}, \citenamefont {Yan}, \citenamefont {Qin},\ and\ \citenamefont {Liu}}]{chen2024AFMspintronics}%
  \BibitemOpen
  \bibfield  {author} {\bibinfo {author} {\bibfnamefont {H.}~\bibnamefont {Chen}}, \bibinfo {author} {\bibfnamefont {L.}~\bibnamefont {Liu}}, \bibinfo {author} {\bibfnamefont {X.}~\bibnamefont {Zhou}}, \bibinfo {author} {\bibfnamefont {Z.}~\bibnamefont {Meng}}, \bibinfo {author} {\bibfnamefont {X.}~\bibnamefont {Wang}}, \bibinfo {author} {\bibfnamefont {Z.}~\bibnamefont {Duan}}, \bibinfo {author} {\bibfnamefont {G.}~\bibnamefont {Zhao}}, \bibinfo {author} {\bibfnamefont {H.}~\bibnamefont {Yan}}, \bibinfo {author} {\bibfnamefont {P.}~\bibnamefont {Qin}},\ and\ \bibinfo {author} {\bibfnamefont {Z.}~\bibnamefont {Liu}},\ }\bibfield  {title} {\bibinfo {title} {Emerging antiferromagnets for spintronics},\ }\href@noop {} {\bibfield  {journal} {\bibinfo  {journal} {Advanced Materials}\ }\textbf {\bibinfo {volume} {36}},\ \bibinfo {pages} {2310379} (\bibinfo {year} {2024})}\BibitemShut {NoStop}%
\bibitem [{\citenamefont {Siddiqui}\ \emph {et~al.}(2020)\citenamefont {Siddiqui}, \citenamefont {Sklenar}, \citenamefont {Kang}, \citenamefont {Gilbert}, \citenamefont {Schleife}, \citenamefont {Mason},\ and\ \citenamefont {Hoffmann}}]{siddiqui2020metallicAFM}%
  \BibitemOpen
  \bibfield  {author} {\bibinfo {author} {\bibfnamefont {S.~A.}\ \bibnamefont {Siddiqui}}, \bibinfo {author} {\bibfnamefont {J.}~\bibnamefont {Sklenar}}, \bibinfo {author} {\bibfnamefont {K.}~\bibnamefont {Kang}}, \bibinfo {author} {\bibfnamefont {M.~J.}\ \bibnamefont {Gilbert}}, \bibinfo {author} {\bibfnamefont {A.}~\bibnamefont {Schleife}}, \bibinfo {author} {\bibfnamefont {N.}~\bibnamefont {Mason}},\ and\ \bibinfo {author} {\bibfnamefont {A.}~\bibnamefont {Hoffmann}},\ }\bibfield  {title} {\bibinfo {title} {Metallic antiferromagnets},\ }\href@noop {} {\bibfield  {journal} {\bibinfo  {journal} {Journal of Applied Physics}\ }\textbf {\bibinfo {volume} {128}} (\bibinfo {year} {2020})}\BibitemShut {NoStop}%
\bibitem [{\citenamefont {Meer}\ \emph {et~al.}(2023)\citenamefont {Meer}, \citenamefont {Gomonay}, \citenamefont {Wittmann},\ and\ \citenamefont {Kl{\"a}ui}}]{meer2023AFMoxides}%
  \BibitemOpen
  \bibfield  {author} {\bibinfo {author} {\bibfnamefont {H.}~\bibnamefont {Meer}}, \bibinfo {author} {\bibfnamefont {O.}~\bibnamefont {Gomonay}}, \bibinfo {author} {\bibfnamefont {A.}~\bibnamefont {Wittmann}},\ and\ \bibinfo {author} {\bibfnamefont {M.}~\bibnamefont {Kl{\"a}ui}},\ }\bibfield  {title} {\bibinfo {title} {Antiferromagnetic insulatronics: Spintronics in insulating 3d metal oxides with antiferromagnetic coupling},\ }\href@noop {} {\bibfield  {journal} {\bibinfo  {journal} {Applied Physics Letters}\ }\textbf {\bibinfo {volume} {122}} (\bibinfo {year} {2023})}\BibitemShut {NoStop}%
\bibitem [{\citenamefont {M{\'a}ca}\ \emph {et~al.}(2012)\citenamefont {M{\'a}ca}, \citenamefont {Ma{\v{s}}ek}, \citenamefont {Stelmakhovych}, \citenamefont {Mart{\'\i}}, \citenamefont {Reichlov{\'a}}, \citenamefont {Uhl{\'\i}{\v{r}}ov{\'a}}, \citenamefont {Beran}, \citenamefont {Wadley}, \citenamefont {Nov{\'a}k},\ and\ \citenamefont {Jungwirth}}]{maca2012CuMnAs}%
  \BibitemOpen
  \bibfield  {author} {\bibinfo {author} {\bibfnamefont {F.}~\bibnamefont {M{\'a}ca}}, \bibinfo {author} {\bibfnamefont {J.}~\bibnamefont {Ma{\v{s}}ek}}, \bibinfo {author} {\bibfnamefont {O.}~\bibnamefont {Stelmakhovych}}, \bibinfo {author} {\bibfnamefont {X.}~\bibnamefont {Mart{\'\i}}}, \bibinfo {author} {\bibfnamefont {H.}~\bibnamefont {Reichlov{\'a}}}, \bibinfo {author} {\bibfnamefont {K.}~\bibnamefont {Uhl{\'\i}{\v{r}}ov{\'a}}}, \bibinfo {author} {\bibfnamefont {P.}~\bibnamefont {Beran}}, \bibinfo {author} {\bibfnamefont {P.}~\bibnamefont {Wadley}}, \bibinfo {author} {\bibfnamefont {V.}~\bibnamefont {Nov{\'a}k}},\ and\ \bibinfo {author} {\bibfnamefont {T.}~\bibnamefont {Jungwirth}},\ }\bibfield  {title} {\bibinfo {title} {Room-temperature antiferromagnetism in cumnas},\ }\href@noop {} {\bibfield  {journal} {\bibinfo  {journal} {Journal of magnetism and magnetic materials}\ }\textbf {\bibinfo {volume} {324}},\ \bibinfo {pages} {1606} (\bibinfo {year} {2012})}\BibitemShut {NoStop}%
\bibitem [{\citenamefont {Ryan}\ \emph {et~al.}(2009)\citenamefont {Ryan}, \citenamefont {Cadogan}, \citenamefont {Ritter}, \citenamefont {Canepa}, \citenamefont {Palenzona},\ and\ \citenamefont {Putti}}]{ryan2009cAFMsuperconductor}%
  \BibitemOpen
  \bibfield  {author} {\bibinfo {author} {\bibfnamefont {D.}~\bibnamefont {Ryan}}, \bibinfo {author} {\bibfnamefont {J.}~\bibnamefont {Cadogan}}, \bibinfo {author} {\bibfnamefont {C.}~\bibnamefont {Ritter}}, \bibinfo {author} {\bibfnamefont {F.}~\bibnamefont {Canepa}}, \bibinfo {author} {\bibfnamefont {A.}~\bibnamefont {Palenzona}},\ and\ \bibinfo {author} {\bibfnamefont {M.}~\bibnamefont {Putti}},\ }\bibfield  {title} {\bibinfo {title} {Coexistence of long-ranged magnetic order and superconductivity in the pnictide superconductor {SmFeAsO1-xFx(x= 0, 0.15)}},\ }\href@noop {} {\bibfield  {journal} {\bibinfo  {journal} {Physical Review B—Condensed Matter and Materials Physics}\ }\textbf {\bibinfo {volume} {80}},\ \bibinfo {pages} {220503} (\bibinfo {year} {2009})}\BibitemShut {NoStop}%
\bibitem [{\citenamefont {Wadley}\ \emph {et~al.}(2016)\citenamefont {Wadley}, \citenamefont {Howells}, \citenamefont {{\v{Z}}elezn{\`y}}, \citenamefont {Andrews}, \citenamefont {Hills}, \citenamefont {Campion}, \citenamefont {Nov{\'a}k}, \citenamefont {Olejn{\'\i}k}, \citenamefont {Maccherozzi}, \citenamefont {Dhesi} \emph {et~al.}}]{wadley2016electrical}%
  \BibitemOpen
  \bibfield  {author} {\bibinfo {author} {\bibfnamefont {P.}~\bibnamefont {Wadley}}, \bibinfo {author} {\bibfnamefont {B.}~\bibnamefont {Howells}}, \bibinfo {author} {\bibfnamefont {J.}~\bibnamefont {{\v{Z}}elezn{\`y}}}, \bibinfo {author} {\bibfnamefont {C.}~\bibnamefont {Andrews}}, \bibinfo {author} {\bibfnamefont {V.}~\bibnamefont {Hills}}, \bibinfo {author} {\bibfnamefont {R.~P.}\ \bibnamefont {Campion}}, \bibinfo {author} {\bibfnamefont {V.}~\bibnamefont {Nov{\'a}k}}, \bibinfo {author} {\bibfnamefont {K.}~\bibnamefont {Olejn{\'\i}k}}, \bibinfo {author} {\bibfnamefont {F.}~\bibnamefont {Maccherozzi}}, \bibinfo {author} {\bibfnamefont {S.}~\bibnamefont {Dhesi}}, \emph {et~al.},\ }\bibfield  {title} {\bibinfo {title} {Electrical switching of an antiferromagnet},\ }\href@noop {} {\bibfield  {journal} {\bibinfo  {journal} {Science}\ }\textbf {\bibinfo {volume} {351}},\ \bibinfo {pages} {587} (\bibinfo {year} {2016})}\BibitemShut {NoStop}%
\bibitem [{\citenamefont {Arpaci}\ \emph {et~al.}(2021)\citenamefont {Arpaci}, \citenamefont {Lopez-Dominguez}, \citenamefont {Shi}, \citenamefont {S{\'a}nchez-Tejerina}, \citenamefont {Garesci}, \citenamefont {Wang}, \citenamefont {Yan}, \citenamefont {Sangwan}, \citenamefont {Grayson}, \citenamefont {Hersam} \emph {et~al.}}]{arpaci2021observation}%
  \BibitemOpen
  \bibfield  {author} {\bibinfo {author} {\bibfnamefont {S.}~\bibnamefont {Arpaci}}, \bibinfo {author} {\bibfnamefont {V.}~\bibnamefont {Lopez-Dominguez}}, \bibinfo {author} {\bibfnamefont {J.}~\bibnamefont {Shi}}, \bibinfo {author} {\bibfnamefont {L.}~\bibnamefont {S{\'a}nchez-Tejerina}}, \bibinfo {author} {\bibfnamefont {F.}~\bibnamefont {Garesci}}, \bibinfo {author} {\bibfnamefont {C.}~\bibnamefont {Wang}}, \bibinfo {author} {\bibfnamefont {X.}~\bibnamefont {Yan}}, \bibinfo {author} {\bibfnamefont {V.~K.}\ \bibnamefont {Sangwan}}, \bibinfo {author} {\bibfnamefont {M.~A.}\ \bibnamefont {Grayson}}, \bibinfo {author} {\bibfnamefont {M.~C.}\ \bibnamefont {Hersam}}, \emph {et~al.},\ }\bibfield  {title} {\bibinfo {title} {Observation of current-induced switching in non-collinear antiferromagnetic irmn3 by differential voltage measurements},\ }\href@noop {} {\bibfield  {journal} {\bibinfo  {journal} {Nature communications}\ }\textbf {\bibinfo {volume} {12}},\ \bibinfo {pages} {3828} (\bibinfo {year}
  {2021})}\BibitemShut {NoStop}%
\bibitem [{\citenamefont {Bodnar}\ \emph {et~al.}(2019)\citenamefont {Bodnar}, \citenamefont {Filianina}, \citenamefont {Bommanaboyena}, \citenamefont {Forrest}, \citenamefont {Maccherozzi}, \citenamefont {Sapozhnik}, \citenamefont {Skourski}, \citenamefont {Kl{\"a}ui},\ and\ \citenamefont {Jourdan}}]{bodnar2019imagingMn2Au}%
  \BibitemOpen
  \bibfield  {author} {\bibinfo {author} {\bibfnamefont {S.~Y.}\ \bibnamefont {Bodnar}}, \bibinfo {author} {\bibfnamefont {M.}~\bibnamefont {Filianina}}, \bibinfo {author} {\bibfnamefont {S.}~\bibnamefont {Bommanaboyena}}, \bibinfo {author} {\bibfnamefont {T.}~\bibnamefont {Forrest}}, \bibinfo {author} {\bibfnamefont {F.}~\bibnamefont {Maccherozzi}}, \bibinfo {author} {\bibfnamefont {A.}~\bibnamefont {Sapozhnik}}, \bibinfo {author} {\bibfnamefont {Y.}~\bibnamefont {Skourski}}, \bibinfo {author} {\bibfnamefont {M.}~\bibnamefont {Kl{\"a}ui}},\ and\ \bibinfo {author} {\bibfnamefont {M.}~\bibnamefont {Jourdan}},\ }\bibfield  {title} {\bibinfo {title} {Imaging of current induced n{\'e}el vector switching in antiferromagnetic {$\text{Mn}_2\text{Au}$}},\ }\href@noop {} {\bibfield  {journal} {\bibinfo  {journal} {Physical Review B}\ }\textbf {\bibinfo {volume} {99}},\ \bibinfo {pages} {140409} (\bibinfo {year} {2019})}\BibitemShut {NoStop}%
\bibitem [{\citenamefont {Amin}\ \emph {et~al.}(2023)\citenamefont {Amin}, \citenamefont {Poole}, \citenamefont {Reimers}, \citenamefont {Barton}, \citenamefont {Dal~Din}, \citenamefont {Maccherozzi}, \citenamefont {Dhesi}, \citenamefont {Nov{\'a}k}, \citenamefont {Krizek}, \citenamefont {Chauhan} \emph {et~al.}}]{amin2023merons}%
  \BibitemOpen
  \bibfield  {author} {\bibinfo {author} {\bibfnamefont {O.}~\bibnamefont {Amin}}, \bibinfo {author} {\bibfnamefont {S.}~\bibnamefont {Poole}}, \bibinfo {author} {\bibfnamefont {S.}~\bibnamefont {Reimers}}, \bibinfo {author} {\bibfnamefont {L.}~\bibnamefont {Barton}}, \bibinfo {author} {\bibfnamefont {A.}~\bibnamefont {Dal~Din}}, \bibinfo {author} {\bibfnamefont {F.}~\bibnamefont {Maccherozzi}}, \bibinfo {author} {\bibfnamefont {S.}~\bibnamefont {Dhesi}}, \bibinfo {author} {\bibfnamefont {V.}~\bibnamefont {Nov{\'a}k}}, \bibinfo {author} {\bibfnamefont {F.}~\bibnamefont {Krizek}}, \bibinfo {author} {\bibfnamefont {J.}~\bibnamefont {Chauhan}}, \emph {et~al.},\ }\bibfield  {title} {\bibinfo {title} {Antiferromagnetic half-skyrmions electrically generated and controlled at room temperature},\ }\href@noop {} {\bibfield  {journal} {\bibinfo  {journal} {Nature Nanotechnology}\ }\textbf {\bibinfo {volume} {18}},\ \bibinfo {pages} {849} (\bibinfo {year} {2023})}\BibitemShut {NoStop}%
\bibitem [{\citenamefont {Kriegner}\ \emph {et~al.}(2016)\citenamefont {Kriegner}, \citenamefont {V{\`y}born{\`y}}, \citenamefont {Olejn{\'\i}k}, \citenamefont {Reichlov{\'a}}, \citenamefont {Nov{\'a}k}, \citenamefont {Marti}, \citenamefont {Gazquez}, \citenamefont {Saidl}, \citenamefont {N{\v{e}}mec}, \citenamefont {Volobuev} \emph {et~al.}}]{kriegner2016AMR}%
  \BibitemOpen
  \bibfield  {author} {\bibinfo {author} {\bibfnamefont {D.}~\bibnamefont {Kriegner}}, \bibinfo {author} {\bibfnamefont {K.}~\bibnamefont {V{\`y}born{\`y}}}, \bibinfo {author} {\bibfnamefont {K.}~\bibnamefont {Olejn{\'\i}k}}, \bibinfo {author} {\bibfnamefont {H.}~\bibnamefont {Reichlov{\'a}}}, \bibinfo {author} {\bibfnamefont {V.}~\bibnamefont {Nov{\'a}k}}, \bibinfo {author} {\bibfnamefont {X.}~\bibnamefont {Marti}}, \bibinfo {author} {\bibfnamefont {J.}~\bibnamefont {Gazquez}}, \bibinfo {author} {\bibfnamefont {V.}~\bibnamefont {Saidl}}, \bibinfo {author} {\bibfnamefont {P.}~\bibnamefont {N{\v{e}}mec}}, \bibinfo {author} {\bibfnamefont {V.}~\bibnamefont {Volobuev}}, \emph {et~al.},\ }\bibfield  {title} {\bibinfo {title} {Multiple-stable anisotropic magnetoresistance memory in antiferromagnetic mnte},\ }\href@noop {} {\bibfield  {journal} {\bibinfo  {journal} {Nature communications}\ }\textbf {\bibinfo {volume} {7}},\ \bibinfo {pages} {11623} (\bibinfo {year} {2016})}\BibitemShut {NoStop}%
\bibitem [{\citenamefont {Qin}\ \emph {et~al.}(2023)\citenamefont {Qin}, \citenamefont {Yan}, \citenamefont {Wang}, \citenamefont {Chen}, \citenamefont {Meng}, \citenamefont {Dong}, \citenamefont {Zhu}, \citenamefont {Cai}, \citenamefont {Feng}, \citenamefont {Zhou} \emph {et~al.}}]{qin2023roomTMR}%
  \BibitemOpen
  \bibfield  {author} {\bibinfo {author} {\bibfnamefont {P.}~\bibnamefont {Qin}}, \bibinfo {author} {\bibfnamefont {H.}~\bibnamefont {Yan}}, \bibinfo {author} {\bibfnamefont {X.}~\bibnamefont {Wang}}, \bibinfo {author} {\bibfnamefont {H.}~\bibnamefont {Chen}}, \bibinfo {author} {\bibfnamefont {Z.}~\bibnamefont {Meng}}, \bibinfo {author} {\bibfnamefont {J.}~\bibnamefont {Dong}}, \bibinfo {author} {\bibfnamefont {M.}~\bibnamefont {Zhu}}, \bibinfo {author} {\bibfnamefont {J.}~\bibnamefont {Cai}}, \bibinfo {author} {\bibfnamefont {Z.}~\bibnamefont {Feng}}, \bibinfo {author} {\bibfnamefont {X.}~\bibnamefont {Zhou}}, \emph {et~al.},\ }\bibfield  {title} {\bibinfo {title} {Room-temperature magnetoresistance in an all-antiferromagnetic tunnel junction},\ }\href@noop {} {\bibfield  {journal} {\bibinfo  {journal} {Nature}\ }\textbf {\bibinfo {volume} {613}},\ \bibinfo {pages} {485} (\bibinfo {year} {2023})}\BibitemShut {NoStop}%
\bibitem [{\citenamefont {Grzybowski}\ \emph {et~al.}(2017)\citenamefont {Grzybowski}, \citenamefont {Wadley}, \citenamefont {Edmonds}, \citenamefont {Beardsley}, \citenamefont {Hills}, \citenamefont {Campion}, \citenamefont {Gallagher}, \citenamefont {Chauhan}, \citenamefont {Novak}, \citenamefont {Jungwirth} \emph {et~al.}}]{grzybowski2017AFimagingPEEM}%
  \BibitemOpen
  \bibfield  {author} {\bibinfo {author} {\bibfnamefont {M.}~\bibnamefont {Grzybowski}}, \bibinfo {author} {\bibfnamefont {P.}~\bibnamefont {Wadley}}, \bibinfo {author} {\bibfnamefont {K.}~\bibnamefont {Edmonds}}, \bibinfo {author} {\bibfnamefont {R.}~\bibnamefont {Beardsley}}, \bibinfo {author} {\bibfnamefont {V.}~\bibnamefont {Hills}}, \bibinfo {author} {\bibfnamefont {R.}~\bibnamefont {Campion}}, \bibinfo {author} {\bibfnamefont {B.}~\bibnamefont {Gallagher}}, \bibinfo {author} {\bibfnamefont {J.~S.}\ \bibnamefont {Chauhan}}, \bibinfo {author} {\bibfnamefont {V.}~\bibnamefont {Novak}}, \bibinfo {author} {\bibfnamefont {T.}~\bibnamefont {Jungwirth}}, \emph {et~al.},\ }\bibfield  {title} {\bibinfo {title} {Imaging current-induced switching of antiferromagnetic domains in cumnas},\ }\href@noop {} {\bibfield  {journal} {\bibinfo  {journal} {Physical review letters}\ }\textbf {\bibinfo {volume} {118}},\ \bibinfo {pages} {057701} (\bibinfo {year} {2017})}\BibitemShut {NoStop}%
\bibitem [{\citenamefont {Sapozhnik}\ \emph {et~al.}(2018{\natexlab{a}})\citenamefont {Sapozhnik}, \citenamefont {Filianina}, \citenamefont {Bodnar}, \citenamefont {Lamirand}, \citenamefont {Mawass}, \citenamefont {Skourski}, \citenamefont {Elmers}, \citenamefont {Zabel}, \citenamefont {Kl{\"a}ui},\ and\ \citenamefont {Jourdan}}]{sapozhnik2018dMn2AuPEEM}%
  \BibitemOpen
  \bibfield  {author} {\bibinfo {author} {\bibfnamefont {A.}~\bibnamefont {Sapozhnik}}, \bibinfo {author} {\bibfnamefont {M.}~\bibnamefont {Filianina}}, \bibinfo {author} {\bibfnamefont {S.~Y.}\ \bibnamefont {Bodnar}}, \bibinfo {author} {\bibfnamefont {A.}~\bibnamefont {Lamirand}}, \bibinfo {author} {\bibfnamefont {M.-A.}\ \bibnamefont {Mawass}}, \bibinfo {author} {\bibfnamefont {Y.}~\bibnamefont {Skourski}}, \bibinfo {author} {\bibfnamefont {H.-J.}\ \bibnamefont {Elmers}}, \bibinfo {author} {\bibfnamefont {H.}~\bibnamefont {Zabel}}, \bibinfo {author} {\bibfnamefont {M.}~\bibnamefont {Kl{\"a}ui}},\ and\ \bibinfo {author} {\bibfnamefont {M.}~\bibnamefont {Jourdan}},\ }\bibfield  {title} {\bibinfo {title} {Direct imaging of antiferromagnetic domains in {$\text{Mn}_2\text{Au}$} manipulated by high magnetic fields},\ }\href@noop {} {\bibfield  {journal} {\bibinfo  {journal} {Physical Review B}\ }\textbf {\bibinfo {volume} {97}},\ \bibinfo {pages} {134429} (\bibinfo {year} {2018}{\natexlab{a}})}\BibitemShut
  {NoStop}%
\bibitem [{\citenamefont {{\v{S}}mejkal}\ \emph {et~al.}(2022)\citenamefont {{\v{S}}mejkal}, \citenamefont {Sinova},\ and\ \citenamefont {Jungwirth}}]{vsmejkal2022altermagnets}%
  \BibitemOpen
  \bibfield  {author} {\bibinfo {author} {\bibfnamefont {L.}~\bibnamefont {{\v{S}}mejkal}}, \bibinfo {author} {\bibfnamefont {J.}~\bibnamefont {Sinova}},\ and\ \bibinfo {author} {\bibfnamefont {T.}~\bibnamefont {Jungwirth}},\ }\bibfield  {title} {\bibinfo {title} {Emerging research landscape of altermagnetism},\ }\href@noop {} {\bibfield  {journal} {\bibinfo  {journal} {Physical Review X}\ }\textbf {\bibinfo {volume} {12}},\ \bibinfo {pages} {040501} (\bibinfo {year} {2022})}\BibitemShut {NoStop}%
\bibitem [{\citenamefont {Kuiper}\ \emph {et~al.}(1993)\citenamefont {Kuiper}, \citenamefont {Searle}, \citenamefont {Rudolf}, \citenamefont {Tjeng},\ and\ \citenamefont {Chen}}]{kuiper1993XMLDantiferro}%
  \BibitemOpen
  \bibfield  {author} {\bibinfo {author} {\bibfnamefont {P.}~\bibnamefont {Kuiper}}, \bibinfo {author} {\bibfnamefont {B.~G.}\ \bibnamefont {Searle}}, \bibinfo {author} {\bibfnamefont {P.}~\bibnamefont {Rudolf}}, \bibinfo {author} {\bibfnamefont {L.}~\bibnamefont {Tjeng}},\ and\ \bibinfo {author} {\bibfnamefont {C.}~\bibnamefont {Chen}},\ }\bibfield  {title} {\bibinfo {title} {X-ray magnetic dichroism of antiferromagnet fe 2 o 3: the orientation of magnetic moments observed by fe 2p x-ray absorption spectroscopy},\ }\href@noop {} {\bibfield  {journal} {\bibinfo  {journal} {Physical review letters}\ }\textbf {\bibinfo {volume} {70}},\ \bibinfo {pages} {1549} (\bibinfo {year} {1993})}\BibitemShut {NoStop}%
\bibitem [{\citenamefont {St\"ohr}\ \emph {et~al.}(1999)\citenamefont {St\"ohr}, \citenamefont {Scholl}, \citenamefont {Regan}, \citenamefont {Anders}, \citenamefont {L\"uning}, \citenamefont {Scheinfein}, \citenamefont {Padmore},\ and\ \citenamefont {White}}]{stohr1999}%
  \BibitemOpen
  \bibfield  {author} {\bibinfo {author} {\bibfnamefont {J.}~\bibnamefont {St\"ohr}}, \bibinfo {author} {\bibfnamefont {A.}~\bibnamefont {Scholl}}, \bibinfo {author} {\bibfnamefont {T.~J.}\ \bibnamefont {Regan}}, \bibinfo {author} {\bibfnamefont {S.}~\bibnamefont {Anders}}, \bibinfo {author} {\bibfnamefont {J.}~\bibnamefont {L\"uning}}, \bibinfo {author} {\bibfnamefont {M.~R.}\ \bibnamefont {Scheinfein}}, \bibinfo {author} {\bibfnamefont {H.~A.}\ \bibnamefont {Padmore}},\ and\ \bibinfo {author} {\bibfnamefont {R.~L.}\ \bibnamefont {White}},\ }\bibfield  {title} {\bibinfo {title} {Images of the antiferromagnetic structure of a nio(100) surface by means of x-ray magnetic linear dichroism spectromicroscopy},\ }\href {https://doi.org/10.1103/PhysRevLett.83.1862} {\bibfield  {journal} {\bibinfo  {journal} {Phys. Rev. Lett.}\ }\textbf {\bibinfo {volume} {83}},\ \bibinfo {pages} {1862} (\bibinfo {year} {1999})}\BibitemShut {NoStop}%
\bibitem [{\citenamefont {St{\"o}hr}\ and\ \citenamefont {Siegmann}(2006)}]{stohr2006magnetism}%
  \BibitemOpen
  \bibfield  {author} {\bibinfo {author} {\bibfnamefont {J.}~\bibnamefont {St{\"o}hr}}\ and\ \bibinfo {author} {\bibfnamefont {H.~C.}\ \bibnamefont {Siegmann}},\ }\bibfield  {title} {\bibinfo {title} {Magnetism},\ }\href@noop {} {\bibfield  {journal} {\bibinfo  {journal} {Solid-State Sciences. Springer, Berlin, Heidelberg}\ }\textbf {\bibinfo {volume} {5}},\ \bibinfo {pages} {236} (\bibinfo {year} {2006})}\BibitemShut {NoStop}%
\bibitem [{\citenamefont {Kune{\v{s}}}\ and\ \citenamefont {Oppeneer}(2003)}]{kunevs2003XMLD}%
  \BibitemOpen
  \bibfield  {author} {\bibinfo {author} {\bibfnamefont {J.}~\bibnamefont {Kune{\v{s}}}}\ and\ \bibinfo {author} {\bibfnamefont {P.~M.}\ \bibnamefont {Oppeneer}},\ }\bibfield  {title} {\bibinfo {title} {Anisotropic x-ray magnetic linear dichroism at the l 2, 3 edges of cubic fe, co, and ni: Ab initio calculations and model theory},\ }\href@noop {} {\bibfield  {journal} {\bibinfo  {journal} {Physical Review B}\ }\textbf {\bibinfo {volume} {67}},\ \bibinfo {pages} {024431} (\bibinfo {year} {2003})}\BibitemShut {NoStop}%
\bibitem [{\citenamefont {Kune{\v{s}}}\ \emph {et~al.}(2004)\citenamefont {Kune{\v{s}}}, \citenamefont {Oppeneer}, \citenamefont {Valencia}, \citenamefont {Abramsohn}, \citenamefont {Mertins}, \citenamefont {Gudat}, \citenamefont {Hecker},\ and\ \citenamefont {Schneider}}]{kunevs2004XMLDPy}%
  \BibitemOpen
  \bibfield  {author} {\bibinfo {author} {\bibfnamefont {J.}~\bibnamefont {Kune{\v{s}}}}, \bibinfo {author} {\bibfnamefont {P.~M.}\ \bibnamefont {Oppeneer}}, \bibinfo {author} {\bibfnamefont {S.}~\bibnamefont {Valencia}}, \bibinfo {author} {\bibfnamefont {D.}~\bibnamefont {Abramsohn}}, \bibinfo {author} {\bibfnamefont {H.-C.}\ \bibnamefont {Mertins}}, \bibinfo {author} {\bibfnamefont {W.}~\bibnamefont {Gudat}}, \bibinfo {author} {\bibfnamefont {M.}~\bibnamefont {Hecker}},\ and\ \bibinfo {author} {\bibfnamefont {C.}~\bibnamefont {Schneider}},\ }\bibfield  {title} {\bibinfo {title} {Understanding the xmld and its magnetocrystalline anisotropy at the {$L_2$}, 3-edges of 3d transition metals},\ }\href@noop {} {\bibfield  {journal} {\bibinfo  {journal} {Journal of magnetism and magnetic materials}\ }\textbf {\bibinfo {volume} {272}},\ \bibinfo {pages} {2146} (\bibinfo {year} {2004})}\BibitemShut {NoStop}%
\bibitem [{\citenamefont {Arenholz}\ \emph {et~al.}(2006)\citenamefont {Arenholz}, \citenamefont {van~der Laan}, \citenamefont {Chopdekar},\ and\ \citenamefont {Suzuki}}]{arenholz2006XMLD}%
  \BibitemOpen
  \bibfield  {author} {\bibinfo {author} {\bibfnamefont {E.}~\bibnamefont {Arenholz}}, \bibinfo {author} {\bibfnamefont {G.}~\bibnamefont {van~der Laan}}, \bibinfo {author} {\bibfnamefont {R.~V.}\ \bibnamefont {Chopdekar}},\ and\ \bibinfo {author} {\bibfnamefont {Y.}~\bibnamefont {Suzuki}},\ }\bibfield  {title} {\bibinfo {title} {Anisotropic x-ray magnetic linear dichroism at the fe l 2, 3 edges in fe 3 o 4},\ }\href@noop {} {\bibfield  {journal} {\bibinfo  {journal} {Physical Review B—Condensed Matter and Materials Physics}\ }\textbf {\bibinfo {volume} {74}},\ \bibinfo {pages} {094407} (\bibinfo {year} {2006})}\BibitemShut {NoStop}%
\bibitem [{\citenamefont {Sapozhnik}\ \emph {et~al.}(2018{\natexlab{b}})\citenamefont {Sapozhnik}, \citenamefont {Filianina}, \citenamefont {Bodnar}, \citenamefont {Lamirand}, \citenamefont {Mawass}, \citenamefont {Skourski}, \citenamefont {Elmers}, \citenamefont {Zabel}, \citenamefont {Kl{\"a}ui},\ and\ \citenamefont {Jourdan}}]{sapozhnik2018afmPEEM}%
  \BibitemOpen
  \bibfield  {author} {\bibinfo {author} {\bibfnamefont {A.}~\bibnamefont {Sapozhnik}}, \bibinfo {author} {\bibfnamefont {M.}~\bibnamefont {Filianina}}, \bibinfo {author} {\bibfnamefont {S.~Y.}\ \bibnamefont {Bodnar}}, \bibinfo {author} {\bibfnamefont {A.}~\bibnamefont {Lamirand}}, \bibinfo {author} {\bibfnamefont {M.-A.}\ \bibnamefont {Mawass}}, \bibinfo {author} {\bibfnamefont {Y.}~\bibnamefont {Skourski}}, \bibinfo {author} {\bibfnamefont {H.-J.}\ \bibnamefont {Elmers}}, \bibinfo {author} {\bibfnamefont {H.}~\bibnamefont {Zabel}}, \bibinfo {author} {\bibfnamefont {M.}~\bibnamefont {Kl{\"a}ui}},\ and\ \bibinfo {author} {\bibfnamefont {M.}~\bibnamefont {Jourdan}},\ }\bibfield  {title} {\bibinfo {title} {Direct imaging of antiferromagnetic domains in {$\text{Mn}_2\text{Au}$} manipulated by high magnetic fields},\ }\href@noop {} {\bibfield  {journal} {\bibinfo  {journal} {Physical Review B}\ }\textbf {\bibinfo {volume} {97}},\ \bibinfo {pages} {134429} (\bibinfo {year} {2018}{\natexlab{b}})}\BibitemShut
  {NoStop}%
\bibitem [{\citenamefont {Lee}\ \emph {et~al.}(2024)\citenamefont {Lee}, \citenamefont {Kim}, \citenamefont {Son}, \citenamefont {Cui}, \citenamefont {Park}, \citenamefont {Zhang}, \citenamefont {Oh}, \citenamefont {Cheong}, \citenamefont {Kleibert},\ and\ \citenamefont {Park}}]{lee2024}%
  \BibitemOpen
  \bibfield  {author} {\bibinfo {author} {\bibfnamefont {Y.}~\bibnamefont {Lee}}, \bibinfo {author} {\bibfnamefont {C.}~\bibnamefont {Kim}}, \bibinfo {author} {\bibfnamefont {S.}~\bibnamefont {Son}}, \bibinfo {author} {\bibfnamefont {J.}~\bibnamefont {Cui}}, \bibinfo {author} {\bibfnamefont {G.}~\bibnamefont {Park}}, \bibinfo {author} {\bibfnamefont {K.-X.}\ \bibnamefont {Zhang}}, \bibinfo {author} {\bibfnamefont {S.}~\bibnamefont {Oh}}, \bibinfo {author} {\bibfnamefont {H.}~\bibnamefont {Cheong}}, \bibinfo {author} {\bibfnamefont {A.}~\bibnamefont {Kleibert}},\ and\ \bibinfo {author} {\bibfnamefont {J.-G.}\ \bibnamefont {Park}},\ }\bibfield  {title} {\bibinfo {title} {Imaging thermally fluctuating n{\'e}el vectors in van der waals antiferromagnet {$\text{Ni}\text{PS}_3$}},\ }\href@noop {} {\bibfield  {journal} {\bibinfo  {journal} {Nano Letters}\ } (\bibinfo {year} {2024})}\BibitemShut {NoStop}%
\bibitem [{\citenamefont {Arai}\ \emph {et~al.}(2012)\citenamefont {Arai}, \citenamefont {Okuda}, \citenamefont {Tanaka}, \citenamefont {Kotsugi}, \citenamefont {Fukumoto}, \citenamefont {Ohkochi}, \citenamefont {Nakamura}, \citenamefont {Matsushita}, \citenamefont {Muro}, \citenamefont {Oura}, \citenamefont {Senba}, \citenamefont {Ohashi}, \citenamefont {Kakizaki}, \citenamefont {Mitsumata},\ and\ \citenamefont {Kinoshita}}]{Arai2012}%
  \BibitemOpen
  \bibfield  {author} {\bibinfo {author} {\bibfnamefont {K.}~\bibnamefont {Arai}}, \bibinfo {author} {\bibfnamefont {T.}~\bibnamefont {Okuda}}, \bibinfo {author} {\bibfnamefont {A.}~\bibnamefont {Tanaka}}, \bibinfo {author} {\bibfnamefont {M.}~\bibnamefont {Kotsugi}}, \bibinfo {author} {\bibfnamefont {K.}~\bibnamefont {Fukumoto}}, \bibinfo {author} {\bibfnamefont {T.}~\bibnamefont {Ohkochi}}, \bibinfo {author} {\bibfnamefont {T.}~\bibnamefont {Nakamura}}, \bibinfo {author} {\bibfnamefont {T.}~\bibnamefont {Matsushita}}, \bibinfo {author} {\bibfnamefont {T.}~\bibnamefont {Muro}}, \bibinfo {author} {\bibfnamefont {M.}~\bibnamefont {Oura}}, \bibinfo {author} {\bibfnamefont {Y.}~\bibnamefont {Senba}}, \bibinfo {author} {\bibfnamefont {H.}~\bibnamefont {Ohashi}}, \bibinfo {author} {\bibfnamefont {A.}~\bibnamefont {Kakizaki}}, \bibinfo {author} {\bibfnamefont {C.}~\bibnamefont {Mitsumata}},\ and\ \bibinfo {author} {\bibfnamefont {T.}~\bibnamefont {Kinoshita}},\ }\bibfield  {title} {\bibinfo {title}
  {Three-dimensional spin orientation in antiferromagnetic domain walls of nio studied by x-ray magnetic linear dichroism photoemission electron microscopy},\ }\href {https://doi.org/10.1103/PhysRevB.85.104418} {\bibfield  {journal} {\bibinfo  {journal} {Phys. Rev. B}\ }\textbf {\bibinfo {volume} {85}},\ \bibinfo {pages} {104418} (\bibinfo {year} {2012})}\BibitemShut {NoStop}%
\bibitem [{\citenamefont {Luo}\ \emph {et~al.}(2023)\citenamefont {Luo}, \citenamefont {Chen}, \citenamefont {Ukleev}, \citenamefont {Wintz}, \citenamefont {Weigand}, \citenamefont {Abrudan}, \citenamefont {Proke{\v{s}}},\ and\ \citenamefont {Radu}}]{luo2023}%
  \BibitemOpen
  \bibfield  {author} {\bibinfo {author} {\bibfnamefont {C.}~\bibnamefont {Luo}}, \bibinfo {author} {\bibfnamefont {K.}~\bibnamefont {Chen}}, \bibinfo {author} {\bibfnamefont {V.}~\bibnamefont {Ukleev}}, \bibinfo {author} {\bibfnamefont {S.}~\bibnamefont {Wintz}}, \bibinfo {author} {\bibfnamefont {M.}~\bibnamefont {Weigand}}, \bibinfo {author} {\bibfnamefont {R.-M.}\ \bibnamefont {Abrudan}}, \bibinfo {author} {\bibfnamefont {K.}~\bibnamefont {Proke{\v{s}}}},\ and\ \bibinfo {author} {\bibfnamefont {F.}~\bibnamefont {Radu}},\ }\bibfield  {title} {\bibinfo {title} {Direct observation of n{\'e}el-type skyrmions and domain walls in a ferrimagnetic {$\text{Dy}\text{Co}_3$} thin film},\ }\href@noop {} {\bibfield  {journal} {\bibinfo  {journal} {Communications Physics}\ }\textbf {\bibinfo {volume} {6}},\ \bibinfo {pages} {218} (\bibinfo {year} {2023})}\BibitemShut {NoStop}%
\bibitem [{\citenamefont {Harrison}\ \emph {et~al.}(2024)\citenamefont {Harrison}, \citenamefont {Jani}, \citenamefont {Hu}, \citenamefont {Lal}, \citenamefont {Lin}, \citenamefont {Popescu}, \citenamefont {Brown}, \citenamefont {Jaouen}, \citenamefont {Ariando},\ and\ \citenamefont {Radaelli}}]{Harrison2024}%
  \BibitemOpen
  \bibfield  {author} {\bibinfo {author} {\bibfnamefont {J.}~\bibnamefont {Harrison}}, \bibinfo {author} {\bibfnamefont {H.}~\bibnamefont {Jani}}, \bibinfo {author} {\bibfnamefont {J.}~\bibnamefont {Hu}}, \bibinfo {author} {\bibfnamefont {M.}~\bibnamefont {Lal}}, \bibinfo {author} {\bibfnamefont {J.-C.}\ \bibnamefont {Lin}}, \bibinfo {author} {\bibfnamefont {H.}~\bibnamefont {Popescu}}, \bibinfo {author} {\bibfnamefont {J.}~\bibnamefont {Brown}}, \bibinfo {author} {\bibfnamefont {N.}~\bibnamefont {Jaouen}}, \bibinfo {author} {\bibfnamefont {A.}~\bibnamefont {Ariando}},\ and\ \bibinfo {author} {\bibfnamefont {P.~G.}\ \bibnamefont {Radaelli}},\ }\bibfield  {title} {\bibinfo {title} {Holographic imaging of antiferromagnetic domains with in-situ magnetic field},\ }\href {https://doi.org/10.1364/OE.508005} {\bibfield  {journal} {\bibinfo  {journal} {Opt. Express}\ }\textbf {\bibinfo {volume} {32}},\ \bibinfo {pages} {5885} (\bibinfo {year} {2024})}\BibitemShut {NoStop}%
\bibitem [{\citenamefont {Shapiro}\ \emph {et~al.}(2014)\citenamefont {Shapiro}, \citenamefont {Yu}, \citenamefont {Tyliszczak}, \citenamefont {Cabana}, \citenamefont {Celestre}, \citenamefont {Chao}, \citenamefont {Kaznatcheev}, \citenamefont {Kilcoyne}, \citenamefont {Maia}, \citenamefont {Marchesini} \emph {et~al.}}]{shapiro2014}%
  \BibitemOpen
  \bibfield  {author} {\bibinfo {author} {\bibfnamefont {D.~A.}\ \bibnamefont {Shapiro}}, \bibinfo {author} {\bibfnamefont {Y.-S.}\ \bibnamefont {Yu}}, \bibinfo {author} {\bibfnamefont {T.}~\bibnamefont {Tyliszczak}}, \bibinfo {author} {\bibfnamefont {J.}~\bibnamefont {Cabana}}, \bibinfo {author} {\bibfnamefont {R.}~\bibnamefont {Celestre}}, \bibinfo {author} {\bibfnamefont {W.}~\bibnamefont {Chao}}, \bibinfo {author} {\bibfnamefont {K.}~\bibnamefont {Kaznatcheev}}, \bibinfo {author} {\bibfnamefont {A.~D.}\ \bibnamefont {Kilcoyne}}, \bibinfo {author} {\bibfnamefont {F.}~\bibnamefont {Maia}}, \bibinfo {author} {\bibfnamefont {S.}~\bibnamefont {Marchesini}}, \emph {et~al.},\ }\bibfield  {title} {\bibinfo {title} {Chemical composition mapping with nanometre resolution by soft x-ray microscopy},\ }\href@noop {} {\bibfield  {journal} {\bibinfo  {journal} {Nature Photonics}\ }\textbf {\bibinfo {volume} {8}},\ \bibinfo {pages} {765} (\bibinfo {year} {2014})}\BibitemShut {NoStop}%
\bibitem [{\citenamefont {Sun}\ \emph {et~al.}(2021)\citenamefont {Sun}, \citenamefont {Sun}, \citenamefont {Yu}, \citenamefont {Mao}, \citenamefont {Tai}, \citenamefont {Zhang}, \citenamefont {Shao}, \citenamefont {Wang}, \citenamefont {Wang},\ and\ \citenamefont {Zhou}}]{Sun2021}%
  \BibitemOpen
  \bibfield  {author} {\bibinfo {author} {\bibfnamefont {T.}~\bibnamefont {Sun}}, \bibinfo {author} {\bibfnamefont {G.}~\bibnamefont {Sun}}, \bibinfo {author} {\bibfnamefont {F.}~\bibnamefont {Yu}}, \bibinfo {author} {\bibfnamefont {Y.}~\bibnamefont {Mao}}, \bibinfo {author} {\bibfnamefont {R.}~\bibnamefont {Tai}}, \bibinfo {author} {\bibfnamefont {X.}~\bibnamefont {Zhang}}, \bibinfo {author} {\bibfnamefont {G.}~\bibnamefont {Shao}}, \bibinfo {author} {\bibfnamefont {Z.}~\bibnamefont {Wang}}, \bibinfo {author} {\bibfnamefont {J.}~\bibnamefont {Wang}},\ and\ \bibinfo {author} {\bibfnamefont {J.}~\bibnamefont {Zhou}},\ }\bibfield  {title} {\bibinfo {title} {Soft x-ray ptychography chemical imaging of degradation in a composite surface-reconstructed {Li}-rich cathode},\ }\href {https://doi.org/10.1021/acsnano.0c08891} {\bibfield  {journal} {\bibinfo  {journal} {ACS Nano}\ }\textbf {\bibinfo {volume} {15}},\ \bibinfo {pages} {1475} (\bibinfo {year} {2021})},\ \bibinfo {note} {pMID: 33356135},\ \Eprint
  {https://arxiv.org/abs/https://doi.org/10.1021/acsnano.0c08891} {https://doi.org/10.1021/acsnano.0c08891} \BibitemShut {NoStop}%
\bibitem [{\citenamefont {Butcher}\ \emph {et~al.}(2024)\citenamefont {Butcher}, \citenamefont {Phillips}, \citenamefont {Chiu}, \citenamefont {Wei}, \citenamefont {Ho}, \citenamefont {Chen}, \citenamefont {Fr{\"o}jdh}, \citenamefont {Baruffaldi}, \citenamefont {Carulla}, \citenamefont {Zhang} \emph {et~al.}}]{butcher2024MultferroicPtycho}%
  \BibitemOpen
  \bibfield  {author} {\bibinfo {author} {\bibfnamefont {T.~A.}\ \bibnamefont {Butcher}}, \bibinfo {author} {\bibfnamefont {N.~W.}\ \bibnamefont {Phillips}}, \bibinfo {author} {\bibfnamefont {C.-C.}\ \bibnamefont {Chiu}}, \bibinfo {author} {\bibfnamefont {C.-C.}\ \bibnamefont {Wei}}, \bibinfo {author} {\bibfnamefont {S.-Z.}\ \bibnamefont {Ho}}, \bibinfo {author} {\bibfnamefont {Y.-C.}\ \bibnamefont {Chen}}, \bibinfo {author} {\bibfnamefont {E.}~\bibnamefont {Fr{\"o}jdh}}, \bibinfo {author} {\bibfnamefont {F.}~\bibnamefont {Baruffaldi}}, \bibinfo {author} {\bibfnamefont {M.}~\bibnamefont {Carulla}}, \bibinfo {author} {\bibfnamefont {J.}~\bibnamefont {Zhang}}, \emph {et~al.},\ }\bibfield  {title} {\bibinfo {title} {Ptychographic nanoscale imaging of the magnetoelectric coupling in freestanding {$\text{BiFeO}_3$}},\ }\href@noop {} {\bibfield  {journal} {\bibinfo  {journal} {Advanced Materials}\ ,\ \bibinfo {pages} {2311157}} (\bibinfo {year} {2024})}\BibitemShut {NoStop}%
\bibitem [{\citenamefont {Donnelly}\ \emph {et~al.}(2016)\citenamefont {Donnelly}, \citenamefont {Scagnoli}, \citenamefont {Guizar-Sicairos}, \citenamefont {Holler}, \citenamefont {Wilhelm}, \citenamefont {Guillou}, \citenamefont {Rogalev}, \citenamefont {Detlefs}, \citenamefont {Menzel}, \citenamefont {Raabe} \emph {et~al.}}]{donnelly2016hardXRAY}%
  \BibitemOpen
  \bibfield  {author} {\bibinfo {author} {\bibfnamefont {C.}~\bibnamefont {Donnelly}}, \bibinfo {author} {\bibfnamefont {V.}~\bibnamefont {Scagnoli}}, \bibinfo {author} {\bibfnamefont {M.}~\bibnamefont {Guizar-Sicairos}}, \bibinfo {author} {\bibfnamefont {M.}~\bibnamefont {Holler}}, \bibinfo {author} {\bibfnamefont {F.}~\bibnamefont {Wilhelm}}, \bibinfo {author} {\bibfnamefont {F.}~\bibnamefont {Guillou}}, \bibinfo {author} {\bibfnamefont {A.}~\bibnamefont {Rogalev}}, \bibinfo {author} {\bibfnamefont {C.}~\bibnamefont {Detlefs}}, \bibinfo {author} {\bibfnamefont {A.}~\bibnamefont {Menzel}}, \bibinfo {author} {\bibfnamefont {J.}~\bibnamefont {Raabe}}, \emph {et~al.},\ }\bibfield  {title} {\bibinfo {title} {High-resolution hard x-ray magnetic imaging with dichroic ptychography},\ }\href@noop {} {\bibfield  {journal} {\bibinfo  {journal} {Physical Review B}\ }\textbf {\bibinfo {volume} {94}},\ \bibinfo {pages} {064421} (\bibinfo {year} {2016})}\BibitemShut {NoStop}%
\bibitem [{\citenamefont {Donnelly}\ \emph {et~al.}(2017)\citenamefont {Donnelly}, \citenamefont {Guizar-Sicairos}, \citenamefont {Scagnoli}, \citenamefont {Gliga}, \citenamefont {Holler}, \citenamefont {Raabe},\ and\ \citenamefont {Heyderman}}]{donnelly2017vectorTOMO}%
  \BibitemOpen
  \bibfield  {author} {\bibinfo {author} {\bibfnamefont {C.}~\bibnamefont {Donnelly}}, \bibinfo {author} {\bibfnamefont {M.}~\bibnamefont {Guizar-Sicairos}}, \bibinfo {author} {\bibfnamefont {V.}~\bibnamefont {Scagnoli}}, \bibinfo {author} {\bibfnamefont {S.}~\bibnamefont {Gliga}}, \bibinfo {author} {\bibfnamefont {M.}~\bibnamefont {Holler}}, \bibinfo {author} {\bibfnamefont {J.}~\bibnamefont {Raabe}},\ and\ \bibinfo {author} {\bibfnamefont {L.~J.}\ \bibnamefont {Heyderman}},\ }\bibfield  {title} {\bibinfo {title} {Three-dimensional magnetization structures revealed with x-ray vector nanotomography},\ }\href@noop {} {\bibfield  {journal} {\bibinfo  {journal} {Nature}\ }\textbf {\bibinfo {volume} {547}},\ \bibinfo {pages} {328} (\bibinfo {year} {2017})}\BibitemShut {NoStop}%
\bibitem [{\citenamefont {Di~Pietro~Mart\'{\i}nez}\ \emph {et~al.}(2025)\citenamefont {Di~Pietro~Mart\'{\i}nez}, \citenamefont {Wartelle}, \citenamefont {Mille}, \citenamefont {Stanescu}, \citenamefont {Belkhou}, \citenamefont {Fettar}, \citenamefont {Favre-Nicolin},\ and\ \citenamefont {Beutier}}]{DiPietro_singlepolPtycho}%
  \BibitemOpen
  \bibfield  {author} {\bibinfo {author} {\bibfnamefont {M.}~\bibnamefont {Di~Pietro~Mart\'{\i}nez}}, \bibinfo {author} {\bibfnamefont {A.}~\bibnamefont {Wartelle}}, \bibinfo {author} {\bibfnamefont {N.}~\bibnamefont {Mille}}, \bibinfo {author} {\bibfnamefont {S.}~\bibnamefont {Stanescu}}, \bibinfo {author} {\bibfnamefont {R.}~\bibnamefont {Belkhou}}, \bibinfo {author} {\bibfnamefont {F.}~\bibnamefont {Fettar}}, \bibinfo {author} {\bibfnamefont {V.}~\bibnamefont {Favre-Nicolin}},\ and\ \bibinfo {author} {\bibfnamefont {G.}~\bibnamefont {Beutier}},\ }\bibfield  {title} {\bibinfo {title} {Magnetic x-ray imaging using a single polarization and multimodal ptychography},\ }\href {https://doi.org/10.1103/PhysRevLett.134.016704} {\bibfield  {journal} {\bibinfo  {journal} {Phys. Rev. Lett.}\ }\textbf {\bibinfo {volume} {134}},\ \bibinfo {pages} {016704} (\bibinfo {year} {2025})}\BibitemShut {NoStop}%
\bibitem [{\citenamefont {Neethirajan}\ \emph {et~al.}(2024)\citenamefont {Neethirajan}, \citenamefont {Daurer}, \citenamefont {Mart\'{\i}nez}, \citenamefont {Hrabec}, \citenamefont {Turnbull}, \citenamefont {Yamamoto}, \citenamefont {Ferreira}, \citenamefont {\ifmmode \check{S}\else \v{S}\fi{}tefan\ifmmode \check{c}\else \v{c}\fi{}i\ifmmode~\check{c}\else \v{c}\fi{}}, \citenamefont {Mayoh}, \citenamefont {Balakrishnan}, \citenamefont {Pei}, \citenamefont {Xue}, \citenamefont {Chang}, \citenamefont {Ringe}, \citenamefont {Harrison}, \citenamefont {Valencia}, \citenamefont {Kazemian}, \citenamefont {Kaulich},\ and\ \citenamefont {Donnelly}}]{Jeffrey2024}%
  \BibitemOpen
  \bibfield  {author} {\bibinfo {author} {\bibfnamefont {J.}~\bibnamefont {Neethirajan}}, \bibinfo {author} {\bibfnamefont {B.~J.}\ \bibnamefont {Daurer}}, \bibinfo {author} {\bibfnamefont {M.~D.~P.}\ \bibnamefont {Mart\'{\i}nez}}, \bibinfo {author} {\bibfnamefont {A.~c.~v.}\ \bibnamefont {Hrabec}}, \bibinfo {author} {\bibfnamefont {L.}~\bibnamefont {Turnbull}}, \bibinfo {author} {\bibfnamefont {R.}~\bibnamefont {Yamamoto}}, \bibinfo {author} {\bibfnamefont {M.~R.}\ \bibnamefont {Ferreira}}, \bibinfo {author} {\bibfnamefont {A.~c.~v.}\ \bibnamefont {\ifmmode \check{S}\else \v{S}\fi{}tefan\ifmmode \check{c}\else \v{c}\fi{}i\ifmmode~\check{c}\else \v{c}\fi{}}}, \bibinfo {author} {\bibfnamefont {D.~A.}\ \bibnamefont {Mayoh}}, \bibinfo {author} {\bibfnamefont {G.}~\bibnamefont {Balakrishnan}}, \bibinfo {author} {\bibfnamefont {Z.}~\bibnamefont {Pei}}, \bibinfo {author} {\bibfnamefont {P.}~\bibnamefont {Xue}}, \bibinfo {author} {\bibfnamefont {L.}~\bibnamefont {Chang}}, \bibinfo {author} {\bibfnamefont
  {E.}~\bibnamefont {Ringe}}, \bibinfo {author} {\bibfnamefont {R.}~\bibnamefont {Harrison}}, \bibinfo {author} {\bibfnamefont {S.}~\bibnamefont {Valencia}}, \bibinfo {author} {\bibfnamefont {M.}~\bibnamefont {Kazemian}}, \bibinfo {author} {\bibfnamefont {B.}~\bibnamefont {Kaulich}},\ and\ \bibinfo {author} {\bibfnamefont {C.}~\bibnamefont {Donnelly}},\ }\bibfield  {title} {\bibinfo {title} {Soft x-ray phase nanomicroscopy of micrometer-thick magnets},\ }\href {https://doi.org/10.1103/PhysRevX.14.031028} {\bibfield  {journal} {\bibinfo  {journal} {Phys. Rev. X}\ }\textbf {\bibinfo {volume} {14}},\ \bibinfo {pages} {031028} (\bibinfo {year} {2024})}\BibitemShut {NoStop}%
\bibitem [{\citenamefont {Francisco}(2022)}]{LUCASphdthesis}%
  \BibitemOpen
  \bibfield  {author} {\bibinfo {author} {\bibfnamefont {L.~H.}\ \bibnamefont {Francisco}},\ }\emph {\bibinfo {title} {X-ray techniques applied to the investigation of superconductors under extreme pressures}},\ \href@noop {} {\bibinfo {type} {Doctoral dissertation}},\ \bibinfo  {school} {State University of Campinas} (\bibinfo {year} {2022})\BibitemShut {NoStop}%
\bibitem [{\citenamefont {Enders}\ and\ \citenamefont {Thibault}(2016)}]{ptypy}%
  \BibitemOpen
  \bibfield  {author} {\bibinfo {author} {\bibfnamefont {B.}~\bibnamefont {Enders}}\ and\ \bibinfo {author} {\bibfnamefont {P.}~\bibnamefont {Thibault}},\ }\bibfield  {title} {\bibinfo {title} {A computational framework for ptychographic reconstructions},\ }\href@noop {} {\bibfield  {journal} {\bibinfo  {journal} {Proceedings of the Royal Society A: Mathematical, Physical and Engineering Sciences}\ }\textbf {\bibinfo {volume} {472}},\ \bibinfo {pages} {20160640} (\bibinfo {year} {2016})}\BibitemShut {NoStop}%
\bibitem [{\citenamefont {Virtanen}\ \emph {et~al.}(2020)\citenamefont {Virtanen}, \citenamefont {Gommers}, \citenamefont {Oliphant}, \citenamefont {Haberland}, \citenamefont {Reddy}, \citenamefont {Cournapeau}, \citenamefont {Burovski}, \citenamefont {Peterson}, \citenamefont {Weckesser}, \citenamefont {Bright}, \citenamefont {{van der Walt}}, \citenamefont {Brett}, \citenamefont {Wilson}, \citenamefont {Millman}, \citenamefont {Mayorov}, \citenamefont {Nelson}, \citenamefont {Jones}, \citenamefont {Kern}, \citenamefont {Larson}, \citenamefont {Carey}, \citenamefont {Polat}, \citenamefont {Feng}, \citenamefont {Moore}, \citenamefont {{VanderPlas}}, \citenamefont {Laxalde}, \citenamefont {Perktold}, \citenamefont {Cimrman}, \citenamefont {Henriksen}, \citenamefont {Quintero}, \citenamefont {Harris}, \citenamefont {Archibald}, \citenamefont {Ribeiro}, \citenamefont {Pedregosa}, \citenamefont {{van Mulbregt}},\ and\ \citenamefont {{SciPy 1.0 Contributors}}}]{2020SciPy}%
  \BibitemOpen
  \bibfield  {author} {\bibinfo {author} {\bibfnamefont {P.}~\bibnamefont {Virtanen}}, \bibinfo {author} {\bibfnamefont {R.}~\bibnamefont {Gommers}}, \bibinfo {author} {\bibfnamefont {T.~E.}\ \bibnamefont {Oliphant}}, \bibinfo {author} {\bibfnamefont {M.}~\bibnamefont {Haberland}}, \bibinfo {author} {\bibfnamefont {T.}~\bibnamefont {Reddy}}, \bibinfo {author} {\bibfnamefont {D.}~\bibnamefont {Cournapeau}}, \bibinfo {author} {\bibfnamefont {E.}~\bibnamefont {Burovski}}, \bibinfo {author} {\bibfnamefont {P.}~\bibnamefont {Peterson}}, \bibinfo {author} {\bibfnamefont {W.}~\bibnamefont {Weckesser}}, \bibinfo {author} {\bibfnamefont {J.}~\bibnamefont {Bright}}, \bibinfo {author} {\bibfnamefont {S.~J.}\ \bibnamefont {{van der Walt}}}, \bibinfo {author} {\bibfnamefont {M.}~\bibnamefont {Brett}}, \bibinfo {author} {\bibfnamefont {J.}~\bibnamefont {Wilson}}, \bibinfo {author} {\bibfnamefont {K.~J.}\ \bibnamefont {Millman}}, \bibinfo {author} {\bibfnamefont {N.}~\bibnamefont {Mayorov}}, \bibinfo {author} {\bibfnamefont
  {A.~R.~J.}\ \bibnamefont {Nelson}}, \bibinfo {author} {\bibfnamefont {E.}~\bibnamefont {Jones}}, \bibinfo {author} {\bibfnamefont {R.}~\bibnamefont {Kern}}, \bibinfo {author} {\bibfnamefont {E.}~\bibnamefont {Larson}}, \bibinfo {author} {\bibfnamefont {C.~J.}\ \bibnamefont {Carey}}, \bibinfo {author} {\bibfnamefont {{\.I}.}~\bibnamefont {Polat}}, \bibinfo {author} {\bibfnamefont {Y.}~\bibnamefont {Feng}}, \bibinfo {author} {\bibfnamefont {E.~W.}\ \bibnamefont {Moore}}, \bibinfo {author} {\bibfnamefont {J.}~\bibnamefont {{VanderPlas}}}, \bibinfo {author} {\bibfnamefont {D.}~\bibnamefont {Laxalde}}, \bibinfo {author} {\bibfnamefont {J.}~\bibnamefont {Perktold}}, \bibinfo {author} {\bibfnamefont {R.}~\bibnamefont {Cimrman}}, \bibinfo {author} {\bibfnamefont {I.}~\bibnamefont {Henriksen}}, \bibinfo {author} {\bibfnamefont {E.~A.}\ \bibnamefont {Quintero}}, \bibinfo {author} {\bibfnamefont {C.~R.}\ \bibnamefont {Harris}}, \bibinfo {author} {\bibfnamefont {A.~M.}\ \bibnamefont {Archibald}}, \bibinfo {author}
  {\bibfnamefont {A.~H.}\ \bibnamefont {Ribeiro}}, \bibinfo {author} {\bibfnamefont {F.}~\bibnamefont {Pedregosa}}, \bibinfo {author} {\bibfnamefont {P.}~\bibnamefont {{van Mulbregt}}},\ and\ \bibinfo {author} {\bibnamefont {{SciPy 1.0 Contributors}}},\ }\bibfield  {title} {\bibinfo {title} {{{SciPy} 1.0: Fundamental Algorithms for Scientific Computing in Python}},\ }\href {https://doi.org/10.1038/s41592-019-0686-2} {\bibfield  {journal} {\bibinfo  {journal} {Nature Methods}\ }\textbf {\bibinfo {volume} {17}},\ \bibinfo {pages} {261} (\bibinfo {year} {2020})}\BibitemShut {NoStop}%
\bibitem [{\citenamefont {Schwickert}\ \emph {et~al.}(1998)\citenamefont {Schwickert}, \citenamefont {Guo}, \citenamefont {Tomaz}, \citenamefont {O’Brien},\ and\ \citenamefont {Harp}}]{schwickert1998FeEdgeXMLD}%
  \BibitemOpen
  \bibfield  {author} {\bibinfo {author} {\bibfnamefont {M.}~\bibnamefont {Schwickert}}, \bibinfo {author} {\bibfnamefont {G.}~\bibnamefont {Guo}}, \bibinfo {author} {\bibfnamefont {M.}~\bibnamefont {Tomaz}}, \bibinfo {author} {\bibfnamefont {W.}~\bibnamefont {O’Brien}},\ and\ \bibinfo {author} {\bibfnamefont {G.}~\bibnamefont {Harp}},\ }\bibfield  {title} {\bibinfo {title} {X-ray magnetic linear dichroism in absorption at the {L} edge of metallic {Co, Fe, Cr, and V}},\ }\href@noop {} {\bibfield  {journal} {\bibinfo  {journal} {Physical Review B}\ }\textbf {\bibinfo {volume} {58}},\ \bibinfo {pages} {R4289} (\bibinfo {year} {1998})}\BibitemShut {NoStop}%
\bibitem [{\citenamefont {Scherz}\ \emph {et~al.}(2007)\citenamefont {Scherz}, \citenamefont {Schlotter}, \citenamefont {Chen}, \citenamefont {Rick}, \citenamefont {St\"ohr}, \citenamefont {L\"uning}, \citenamefont {McNulty}, \citenamefont {G\"unther}, \citenamefont {Radu}, \citenamefont {Eberhardt}, \citenamefont {Hellwig},\ and\ \citenamefont {Eisebitt}}]{Scherz07}%
  \BibitemOpen
  \bibfield  {author} {\bibinfo {author} {\bibfnamefont {A.}~\bibnamefont {Scherz}}, \bibinfo {author} {\bibfnamefont {W.~F.}\ \bibnamefont {Schlotter}}, \bibinfo {author} {\bibfnamefont {K.}~\bibnamefont {Chen}}, \bibinfo {author} {\bibfnamefont {R.}~\bibnamefont {Rick}}, \bibinfo {author} {\bibfnamefont {J.}~\bibnamefont {St\"ohr}}, \bibinfo {author} {\bibfnamefont {J.}~\bibnamefont {L\"uning}}, \bibinfo {author} {\bibfnamefont {I.}~\bibnamefont {McNulty}}, \bibinfo {author} {\bibfnamefont {C.}~\bibnamefont {G\"unther}}, \bibinfo {author} {\bibfnamefont {F.}~\bibnamefont {Radu}}, \bibinfo {author} {\bibfnamefont {W.}~\bibnamefont {Eberhardt}}, \bibinfo {author} {\bibfnamefont {O.}~\bibnamefont {Hellwig}},\ and\ \bibinfo {author} {\bibfnamefont {S.}~\bibnamefont {Eisebitt}},\ }\bibfield  {title} {\bibinfo {title} {Phase imaging of magnetic nanostructures using resonant soft x-ray holography},\ }\href {https://doi.org/10.1103/PhysRevB.76.214410} {\bibfield  {journal} {\bibinfo  {journal} {Phys. Rev. B}\
  }\textbf {\bibinfo {volume} {76}},\ \bibinfo {pages} {214410} (\bibinfo {year} {2007})}\BibitemShut {NoStop}%
\bibitem [{\citenamefont {Amin}\ \emph {et~al.}(2024)\citenamefont {Amin}, \citenamefont {Dal~Din}, \citenamefont {Golias}, \citenamefont {Niu}, \citenamefont {Zakharov}, \citenamefont {Fromage}, \citenamefont {Fields}, \citenamefont {Heywood}, \citenamefont {Cousins}, \citenamefont {Maccherozzi} \emph {et~al.}}]{amin2024altermagnetism}%
  \BibitemOpen
  \bibfield  {author} {\bibinfo {author} {\bibfnamefont {O.}~\bibnamefont {Amin}}, \bibinfo {author} {\bibfnamefont {A.}~\bibnamefont {Dal~Din}}, \bibinfo {author} {\bibfnamefont {E.}~\bibnamefont {Golias}}, \bibinfo {author} {\bibfnamefont {Y.}~\bibnamefont {Niu}}, \bibinfo {author} {\bibfnamefont {A.}~\bibnamefont {Zakharov}}, \bibinfo {author} {\bibfnamefont {S.}~\bibnamefont {Fromage}}, \bibinfo {author} {\bibfnamefont {C.}~\bibnamefont {Fields}}, \bibinfo {author} {\bibfnamefont {S.}~\bibnamefont {Heywood}}, \bibinfo {author} {\bibfnamefont {R.}~\bibnamefont {Cousins}}, \bibinfo {author} {\bibfnamefont {F.}~\bibnamefont {Maccherozzi}}, \emph {et~al.},\ }\bibfield  {title} {\bibinfo {title} {Nanoscale imaging and control of altermagnetism in mnte},\ }\href@noop {} {\bibfield  {journal} {\bibinfo  {journal} {Nature}\ }\textbf {\bibinfo {volume} {636}},\ \bibinfo {pages} {348} (\bibinfo {year} {2024})}\BibitemShut {NoStop}%
\bibitem [{\citenamefont {Apseros}\ \emph {et~al.}(2024)\citenamefont {Apseros}, \citenamefont {Scagnoli}, \citenamefont {Holler}, \citenamefont {Guizar-Sicairos}, \citenamefont {Gao}, \citenamefont {Appel}, \citenamefont {Heyderman}, \citenamefont {Donnelly},\ and\ \citenamefont {Ihli}}]{Apseros2024XMLDtomo}%
  \BibitemOpen
  \bibfield  {author} {\bibinfo {author} {\bibfnamefont {A.}~\bibnamefont {Apseros}}, \bibinfo {author} {\bibfnamefont {V.}~\bibnamefont {Scagnoli}}, \bibinfo {author} {\bibfnamefont {M.}~\bibnamefont {Holler}}, \bibinfo {author} {\bibfnamefont {M.}~\bibnamefont {Guizar-Sicairos}}, \bibinfo {author} {\bibfnamefont {Z.}~\bibnamefont {Gao}}, \bibinfo {author} {\bibfnamefont {C.}~\bibnamefont {Appel}}, \bibinfo {author} {\bibfnamefont {L.~J.}\ \bibnamefont {Heyderman}}, \bibinfo {author} {\bibfnamefont {C.}~\bibnamefont {Donnelly}},\ and\ \bibinfo {author} {\bibfnamefont {J.}~\bibnamefont {Ihli}},\ }\bibfield  {title} {\bibinfo {title} {X-ray linear dichroic tomography of crystallographic and topological defects},\ }\href {https://doi.org/10.1038/s41586-024-08233-y} {\bibfield  {journal} {\bibinfo  {journal} {Nature}\ }\textbf {\bibinfo {volume} {636}},\ \bibinfo {pages} {354} (\bibinfo {year} {2024})}\BibitemShut {NoStop}%
\bibitem [{\citenamefont {Lovesey}\ and\ \citenamefont {Collins}(1996)}]{lovesey1996x}%
  \BibitemOpen
  \bibfield  {author} {\bibinfo {author} {\bibfnamefont {S.~W.}\ \bibnamefont {Lovesey}}\ and\ \bibinfo {author} {\bibfnamefont {S.~P.}\ \bibnamefont {Collins}},\ }\href@noop {} {\emph {\bibinfo {title} {X-ray scattering and absorption by magnetic materials}}}\ (\bibinfo  {publisher} {Oxford university press},\ \bibinfo {year} {1996})\BibitemShut {NoStop}%
\bibitem [{\citenamefont {van~der Laan}(2008)}]{van2008soft}%
  \BibitemOpen
  \bibfield  {author} {\bibinfo {author} {\bibfnamefont {G.}~\bibnamefont {van~der Laan}},\ }\bibfield  {title} {\bibinfo {title} {Soft x-ray resonant magnetic scattering of magnetic nanostructures},\ }\href@noop {} {\bibfield  {journal} {\bibinfo  {journal} {Comptes Rendus Physique}\ }\textbf {\bibinfo {volume} {9}},\ \bibinfo {pages} {570} (\bibinfo {year} {2008})}\BibitemShut {NoStop}%
\bibitem [{\citenamefont {van~der Walt}\ \emph {et~al.}(2014)\citenamefont {van~der Walt}, \citenamefont {{S}ch\"onberger}, \citenamefont {{Nunez-Iglesias}}, \citenamefont {{B}oulogne}, \citenamefont {{W}arner}, \citenamefont {{Y}ager}, \citenamefont {{G}ouillart}, \citenamefont {{Y}u},\ and\ \citenamefont {the scikit-image contributors}}]{scikit-image}%
  \BibitemOpen
  \bibfield  {author} {\bibinfo {author} {\bibfnamefont {S.}~\bibnamefont {van~der Walt}}, \bibinfo {author} {\bibfnamefont {J.~L.}\ \bibnamefont {{S}ch\"onberger}}, \bibinfo {author} {\bibfnamefont {J.}~\bibnamefont {{Nunez-Iglesias}}}, \bibinfo {author} {\bibfnamefont {F.}~\bibnamefont {{B}oulogne}}, \bibinfo {author} {\bibfnamefont {J.~D.}\ \bibnamefont {{W}arner}}, \bibinfo {author} {\bibfnamefont {N.}~\bibnamefont {{Y}ager}}, \bibinfo {author} {\bibfnamefont {E.}~\bibnamefont {{G}ouillart}}, \bibinfo {author} {\bibfnamefont {T.}~\bibnamefont {{Y}u}},\ and\ \bibinfo {author} {\bibnamefont {the scikit-image contributors}},\ }\bibfield  {title} {\bibinfo {title} {scikit-image: image processing in {P}ython},\ }\href {https://doi.org/10.7717/peerj.453} {\bibfield  {journal} {\bibinfo  {journal} {PeerJ}\ }\textbf {\bibinfo {volume} {2}},\ \bibinfo {pages} {e453} (\bibinfo {year} {2014})}\BibitemShut {NoStop}%
\bibitem [{\citenamefont {{van Heel}}\ and\ \citenamefont {Schatz}(2005)}]{VANHEEL2005FRC}%
  \BibitemOpen
  \bibfield  {author} {\bibinfo {author} {\bibfnamefont {M.}~\bibnamefont {{van Heel}}}\ and\ \bibinfo {author} {\bibfnamefont {M.}~\bibnamefont {Schatz}},\ }\bibfield  {title} {\bibinfo {title} {Fourier shell correlation threshold criteria},\ }\href {https://doi.org/https://doi.org/10.1016/j.jsb.2005.05.009} {\bibfield  {journal} {\bibinfo  {journal} {Journal of Structural Biology}\ }\textbf {\bibinfo {volume} {151}},\ \bibinfo {pages} {250} (\bibinfo {year} {2005})}\BibitemShut {NoStop}%
\end{thebibliography}
%

\end{document}